\documentclass[3p,time]{./elsarticle}
\usepackage{./ecrc}
\def\aplt{\ {\raise-.5ex\hbox{$\buildrel<\over\sim$}}\ }

\def\aap{\ {A\&A}\ }

\def\aj{\ {AJ}\ }

\def\apj{\ {ApJ}\ }
\def\apjl{\ {ApJL}\ }

\def\icarus{\ {Icarus}\ }

\def\mnras{\ {MNRAS}\ }

\def\nat{\ {Nat}\ }

\def\na{\ {New Astron.}\ }

\def\pasj{\ {Publ. Astr. Soc. Japan}\ }

\def\prd{\ {Phys. Rev. D.}\ }

\def\rmxaa{\ {Revista Mexicana de Astronomia y Astrofisica}\ }

\newcommand{\MSun}{\mbox{${\rm M}_\odot$}}

\def\lteq{\ {\raise-.5ex\hbox{$\buildrel<\over-$}}\ }
\def\apgt{\ {\raise-.5ex\hbox{$\buildrel>\over\sim$}}\ }
\def\aplt{\ {\raise-.5ex\hbox{$\buildrel<\over\sim$}}\ }
\def\lt{\ {\raise-.5ex\hbox{$\buildrel>$}}\ }
\def\gt{\ {\raise-.5ex\hbox{$\buildrel<$}}\ }
\def\eqgt{\ {\raise-.5ex\hbox{$\buildrel>\over-$}}\ }
\def\eqlt{\ {\raise-.5ex\hbox{$\buildrel<\over-$}}\ }

\newfont{\Giga}{cmssbx10 scaled 5200}
\newfont{\giga}{cmssbx10 scaled 4300}
\newfont{\Mega}{cmssbx10 scaled 3200}
\newfont{\mega}{cmssbx10 scaled 2500}
\newfont{\Kilo}{cmssbx10 scaled 2000}
\newfont{\kilo}{cmssbx10 scaled 1600}
\newfont{\Deca}{cmssbx10 scaled 1450}
\newfont{\deca}{cmssbx10 scaled 1200}
\newfont{\Dezi}{cmssbx10 scaled 1100}
\newfont{\dezi}{cmssbx10 scaled 1050}
\newfont{\iGiga}{cmssi10 scaled 6200}
\newfont{\igiga}{cmssi10 scaled 4300}
\newfont{\iMega}{cmssi10 scaled 3200}
\newfont{\imega}{cmssi10 scaled 2500}
\newfont{\iKilo}{cmssi10 scaled 2000}
\newfont{\ikilo}{cmssi10 scaled 1500}
\newfont{\mathGiga}{cmsy10 scaled 6200}
\newfont{\mathgiga}{cmsy10 scaled 4300}
\newfont{\mathMega}{cmsy10 scaled 3200}
\newfont{\mathmega}{cmsy10 scaled 2500}
\newfont{\mathKilo}{cmsy10 scaled 2000}
\newfont{\mathkilo}{cmsy10 scaled 1500}
\newfont{\mathDeca}{cmsy10 scaled 1450}
\newfont{\mathdeca}{cmsy10 scaled 1200}
\usepackage{amsmath,amssymb}
\usepackage{rotating}
\usepackage{times}
\usepackage{boxedminipage}
\usepackage{graphicx}
\usepackage{ulem}

\usepackage{mathtools} 
\usepackage{extarrows} 

\usepackage{caption,graphicx,newfloat}
\DeclareCaptionType{codesnippet}

\volume{00}
\firstpage{1}
\journalname{Communications in Nonlinear Science and Numerical Simulation}
\runauth{Portegies Zwart et al.}

\def\aplt{\ {\raise-.5ex\hbox{$\buildrel<\over\sim$}}\ }

\newcommand\vect[1]{\ensuremath{\mathbf{#1}}}
\newcommand\bridge{\textsc{Bridge}}
\newcommand\amuse{\textsc{Amuse}}
\newcommand\omuse{\textsc{Omuse}}
\newcommand\nemesis{\textsc{Nemesis}}

\def\apgt{\ {\raise-.5ex\hbox{$\buildrel>\over\sim$}}\ }
\def\aplt{\ {\raise-.5ex\hbox{$\buildrel<\over\sim$}}\ }
\def\lteq{\ {\raise-.5ex\hbox{$\buildrel<\over-$}}\ }

\renewcommand\gets{\ \longleftarrow\ }
\newcommand\lBridge[1]{\ \xleftarrow[#1] \ }
\newcommand\lrBridge[1]{\ \xleftrightarrow[#1]{\hspace*{3mm}}\ }
\newcommand\lRBridge[1]{\ \xleftarrow[#1]{r} \ }

\def\mnras{\ {MNRAS}\ }
\def\nat{\ {Nature}\ }
\def\aj{\ {AJ}\ }
\def\apj{\ {ApJ}\ }

\def\pasj{\ {Publ. Astr. Soc. Japan}\ }

\def\lt{\ {\raise-.5ex\hbox{$<$}}\ }
\def\gt{\ {\raise-.5ex\hbox{$>$}}\ }

\begin{document}
\begin{frontmatter}
  
\dochead{Research}

\title{Non-intrusive hierarchical coupling strategies for multi-scale simulations in gravitational dynamics}

\author{Simon Portegies Zwart$^1$,
  Inti Pelupessy$^2$,
  Carmen Mart\'inez-Barbosa$^3$,
  Arjen van Elteren$^1$,
  Steve McMillan$^5$
  %%Guilherme Gon\c calves Ferrari, %%gg.ferrari@gmail.com
}

\address{
  $^{1}$ Leiden Observatory, Leiden University, PO Box 9513, 2300 RA, Leiden, The Netherlands. \\
  $^{2}$ Netherlands eScience Center, Amsterdam \\
  $^{3}$ Deltares Software Center unit, Delft, The Netherlands.  \\
  %%  E-mail: carmen.martinezbarbosa@deltares.nl \\
  $^{4}$ Drexel University, Department of Physics and Astronomy,
        Disque Hall, 32 S 32nd St., Philadelphia, PA 19104, USA
}

\begin{abstract} % abstract
Hierarchical code coupling strategies make it possible to combine the
results of individual numerical solvers into a self-consistent
symplectic solution. We explore the possibility of allowing such a
coupling strategy to be non-intrusive. In that case, the underlying
numerical implementation is not affected by the coupling itself, but
its functionality is carried over in the interface.  This method is
efficient for solving the equations of motion for a self-gravitating
system over a wide range of scales. We adopt a dedicated integrator
for solving each particular part of the problem and combine the
results to a self-consistent solution. In particular, we explore the
possibilities of combining the evolution of one or more microscopic
systems that are embedded in a macroscopic system. The here presented
generalizations of \bridge\ include higher-order coupling strategies
(from the classic $2^\textrm{nd}$ order up to $10^\textrm{th}$-order),
but we also demonstrate how multiple bridges can be nested and how
additional processes can be introduced at the bridge time-step to
enrich the physics, for example by incorporating dissipative
processesor. Such augmentation allows for including additional
processes in a classic Newtonian $N$-body integrator without
alterations to the underlying code. These additional processes include
for example the Yarkovsky effect, dynamical friction or relativistic
dynamics. Some of these processes operate on all particles whereas
others apply only to a subset.

The presented method is non-intrusive in the sense that the underlying
methods remain operational without changes to the code (apart from
adding the get- and set-functions to enable the bridge operator). As
a result, the fundamental integrators continue to operate with their
internal time step and preserve their local optimizations and
parallelism. Multiple bridges can be nested and coupled
hierarchically, allowing for the construction of a complex environment
of multiple nested augmented bridges. While the coupling topology may
become rather complicated, we introduce the hierarchical coupling
language (HCL), a meta language in which complex bridge topologies can
be described. The meta language is meant for stimulating the discussion
on even more complex hierarchies in which the bridge operators are
introduced as patterns

We present example applications for several of these cases and discuss
the conditions under which these integrators can be applied. Typical
applications range over 10 orders of magnitude in temporal and spatial
scales when we apply the method to simulating planetary systems (au
spatial and year-temporal scale) in a star cluster that orbits in the
Galaxy (100\,kpc-spatial and 10\,Gyr-temporal scale).

\end{abstract}

\end{frontmatter}

\section{Introduction}
\label{Sect:introduction}

Scientific progress is mediated by performing computer simulations of
laboratory experiments or observed phenomena. This is generally done
by executing complex instructions on digital computers that are
inspired by a Turing machine \cite{turing1936a}. The results of these
simulations are subsequently interpreted with similar trust as if they
were produced in a lab experiment or from observations
\cite{Oberkampf:2010:VVS:1941955}. The exponential growth in computer
power \cite{Moore} and improved expressiveness in computer languages
\cite{FELLEISEN199135} enables researchers to perform ever more
complex calculations. At the same time, multi-messenger observations
and the gradual increase in the resolution of lab experiments require
simulations to include more details, wider scales and a broader
palette of phenomena and therefore they become more complex.

Simulation research is slowly progressing into a multi-scale regime.
Once the range in scales exceeds three or four orders of magnitude, we
refer to them as multi-scale simulations
\cite{doi:10.1098/rsta.2018.0355}. The macroscopic scale in
multi-scale simulations tend to progress slowly compared to the
microscopic scale and a large number of operations needed in the
latter tends to introduce round-off and convergence errors.

In those cases, one of the extremes, typically the microscopic scale,
is addressed differently than the rest of the system. Speed-up and
improved convergence is then often realized by incorporating
(semi)analytic solutions or approximations. As a
consequence, the underlying code becomes specific for the particular
problem: The scale transition tends to be hard-coded for each
problem-dependent topology.

Problem specific implementations often lead to the boundless growth of
the source code due to adaptations made upon reuse. In such
dinosource \cite{2019Sci...364...66M}, the newly added functionality
does not increase functionality because deprecated
code is not removed but left to decay \cite{Fowler1999}.

Dinosource can be prevented if the individual components of the
multi-scale hybridization would be non-intrusive. This allows
extensions of the original functionality to be added without affecting
any other parts of the code.

If multi-scale problems can be effectively addressed by non-intrusive
code-coupling strategies, it may also be possible to expand the
range of the domain-specific code by coupling it with codes
from a completely different domain. This strategy mediates performing
multi-physics simulations without interfering with the direct
functionality of the individual solvers. Such non-intrusive
domain-specific coupling will not only support multi-scale simulation
environments but also multi-physics.

Non-intrusiveness code-coupling allows code development 
independently from any of the other solvers. This strict separation of
functionality makes it possible to develop a dedicated implementation 
on one domain at some scale without having to worry about any of
the other scales or domains. It makes code confined, clean, readable
and maintainable. Code that solves for the same physics on the same
scale but written for a different architecture may be developed in
parallel and independently of each other.

One of the advantages of this Duplo-approach
\cite{2019Sci...364...66M} is support for additional functionality in
the form of extensions and expansions. In the former, we envision
additional functionality inside the coupling method that adds
functionality to the data-driven domain. Examples include boundary
conditions or run time variations to the coordinate systems. The
latter case includes additions to the underlying physics that operates
on the same temporal or spatial scale.

There is no particular reason why code coupling should be limited to
two scales or domains, but it is possible to build a cascade of
solvers. With such hybridization, one can extend the coupling over many
orders of magnitude with a cascade of dedicated solvers each
addressing a limited range of scales.

Another attractive aspect is the possibility of hybridizing coupling
strategies in a complex hierarchy of codes. If the implementation
supports controlling across-scale and across-disciplinary interactions
it becomes possible to tune the scale or nature of the interactions
depending on the problem.

If truly non-intrusive, the coupling topology may be allowed to change
at run time. If, for example, the bottleneck in the simulation shifts,
because some criterion is satisfied or a local situation emerges, some
codes may be replaced or terminated whereas others may be
initiated. This allows for flexible run-time behavior, at the cost of
transparency for supercomputer job-scheduling brokers
\cite{2008PCAA.book.....H,2011CS&D....4a5001G}.

In the non-intrusive multi-scale and multi-physics coupling strategy
presented here, all these requirements are satisfied. The coupling
strategy is non-intrusive and it can be controlled at run time,
expanded, extended, hybridized and hierarchically nested.

In this paper, we describe the method (see \S\,\ref{Sect:Bridge}), its
adaptations and extensions (see \S\,\ref{Sect:Adaptations}) and we
provide a more practical notation in the hierarchical coupling
language (HLC) in \S\,\ref{sec:notation}). Eventually, in
\S\,\ref{sec:examples} we provide validation and examples. But first
we review the problem in \S\,\ref{Sect:SSM} from the point of view of
gravitational dynamics.

\section{Non-intrusive coupling strategy for gravitational dynamics}\label{Sect:Bridge}

\subsection{Scale separation in gravitational dynamics}\label{Sect:SSM}

Gravitational dynamics provides an excellent starting point because 
the computational complexity of the problem demands novel
software. Newton's \cite{Newton:1687} equations of motion for $N > 2$
self-gravitating mass-points is one of the oldest and most outstanding
problems in astrophysics \cite{valtonen_karttunen_2006}. The lack of
a practical analytic solution together with the intrinsic chaotic
behavior of the system demands numerical integration to extremely
high precision and accuracy \cite{PORTEGIESZWART2018160}. The
unfavorable scaling of the compute time (${\cal O}(N^2)$ for a direct
force-evaluation scheme) requires enormous resources even for a
relatively small problem. To overcome these limitations a wide variety
of algorithms have been designed in which computer time can be reduced
by trading accuracy for speed. In this way, families of algorithms are
dedicated (and considered suitable by the community) for addressing
specific problems in astrophysics.

One of these problems includes the evolution of a star cluster in
orbit around the Galactic center. This problem is too expensive in
terms of computer time to be integrated with an accurate method in
which the forces are evaluated directly, whereas a hierarchical method
such as a tree-code \cite{1986Natur.324..446B} is insufficiently
precise.

This lead Michiko Fujii and co-workers in 2007 to design a strategy in
which both methods could be hybridized \cite{2007PASJ...59.1095F}.
Their method, called \bridge, is based on a second-order extension of
the mixed-variable symplectic scheme developed in the context of long
term integrations of planetary systems \cite{1991AJ....102.1528W}.

The classic implementation is rather rigid but combines a direct
$4^\textrm{th}$-order predictor-corrector Hermite scheme
\citep{1992PASJ...44..141M,1998NewA....3..309M} with a $2^\textrm{nd}$-order
tree-code. The coupling method itself is $2^\textrm{nd}$ order. The coupling
requires the specification of a rigid cross-over time step. The length
of this step is chosen to minimize the error produced in the
interface. Properly choosing this step size requires some knowledge of
the system and assumes that it does not change much with time. This
hybridization allows the integration of a small but dense star cluster
to be carried out with high accuracy together with a large number of
field stars using low accuracy. A similar strategy was later employed
for studying merging supermassive black holes in galactic nuclei
\cite{2015ComAC...2....6I} and to simulate galaxy mergers
\cite{2013MNRAS.431..767B}. We present a generalization of this
\bridge\, method to higher-order and to include dissipative forces but
most importantly to facilitate non-intrusive coupling.

\begin{figure}[h!]
\center
\includegraphics[clip, trim=4cm 11cm 3cm 10cm,width=0.7\textwidth]{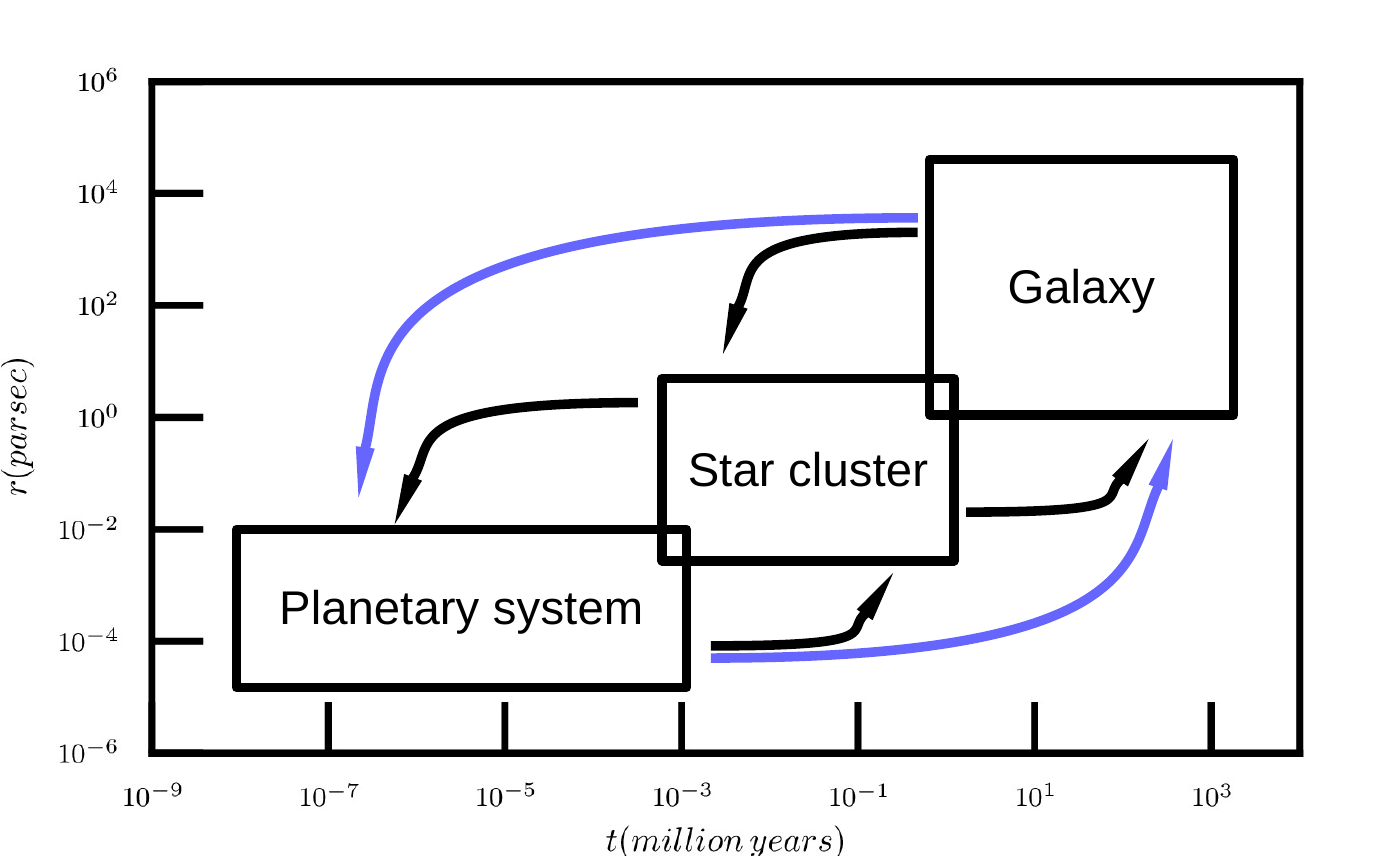}
\caption{Scale Separation Map (SSM) for interaction between planets
and stars in the Galaxy. In planetary systems, interactions take
place on a timescale of days to 1000 years and on spatial scales
ranging from 0.3 to about 100\,au. For the Solar system, these
scales should be associated with the orbit of the planet Mercury
on the small scales and the Kuiper belt on the longest scales.
The stars in a clusters interact on time scales ranging from
$10^3$ to $10^6$ years and typical distance scales range from
$10^4$\,au to about 1\,parsec. Galactic scales range in billion
years and 30\,kilo-parsec. The arrows indicate the interaction
between neighboring systems, whereas the blue arrows show the
cross-scale interactions. The slight overlap of the various
domains illustrate the sometimes fuzzy devision between these, and
the demand for introducing cross-boundary domain solutions. }
\label{fig:ssm_bridge}
\end{figure}

The range in scales of the problem we have in mind is illustrated in a
scale separation map (SSM) \cite{conf.eScience.BorgdorffLHFC11}. In
figure \ref{fig:ssm_bridge}, we present the SSM for planetary systems
(bottom left) in star-clusters (middle) which are a part of the Galaxy
(top right).

The evolution of planetary systems in a star cluster which again is
orbiting the Galaxy provides an excellent example because all aspects
of the non-intrusive hierarchical coupling strategy can be debated in
this context. Planetary systems tend to be composed of one star (or
maybe two) which is orbited by some tens of planets, hundreds of moons
and millions of minor bodies. They are commonly resolved using a
symplectic method in which energy is preserved on a secular time scale
while conserving the orbital phase \cite{1999ssd..book.....M}. Due to
the generally tight orbits of some of the planets or moons,
integration time steps can be as small as days or hours.

The majority of planetary systems are born in clustered environments
with thousands of members. The dynamical time-scales of these systems
range from a few thousand to a few million years. Star clusters
are commonly integrated using the Hermite scheme, which is not
symplectic but $4^\textrm{th}$ or $6^\textrm{th}$ order. Tight planets
may experience thousands of integration steps within a single
integration step of a star in the cluster. A non-intrusive coupling
strategy may be beneficial in this case because it leaves both the
symplectic and the direct Hermite codes unaltered.

The star cluster itself orbits a galaxy. The latter is composed of
hundreds of billion stars and the mean orbital time scale is of the
order of a few hundred million years. The large number of stars in a
galaxy demands a fast integration method with a scaling much better
than $N^2$ and time steps thousands of times larger than those
employed in the star clusters. Galaxies, therefore, tend to be
integrated using a hierarchical method
\cite{1986Natur.324..446B,Bedorf:2014:PGT:2683593.2683600} or a
self-consistent mean-field
\citep{RevModPhys.75.121,2014ApJ...792...98M} or fast multi-pole
\citep{2014CAC.....1....1D} method.

The spatial and temporal separation, illustrated in
Fig.\,\ref{fig:ssm_bridge} shows that these systems tend to be
separated in temporal and spatial scales. In the figure, we introduce
a little overlap to indicate that there may be a mesoscale,
but in most applications, the various domains are well separated.

There is no one-size-fits-all $N$-body solution for the entire domain.
Integrating a galaxy with methods generally used in planetary dynamics
will not give you much progress, whereas the integration of a
planetary system using one of the methods designed for galactic
dynamics will not give you a (community accepted) interpret-able
solution. The common practice is that planets and star clusters are
generally ignored in galactic dynamics, and in studies on planetary dynamics the Galaxy and star clusters are neglected.

This poses no problem so long as planets stay close to their host star
and stars stay in their clusters. In those cases, one could solve the
entire problem as a number of individual disconnected
problems. However, a planet may escape its host star to become a
cluster member or even part of the Galaxy at large, or galactic stars
may be captures by a star cluster and interact with a local planetary
system. In this way, particles can change domain at run time. This is
illustrated with the arrows in Fig.\,\ref{fig:ssm_bridge}. This
exchange of objects, and therefore of information between the
microscopic, the mesoscopic and the macroscopic system demands for a
method in which each of these systems is resolved to sufficient
accuracy and precision.

The strict separation of the particular solvers for the sub-domains
while realizing the coupling non-intrusively has several
advantages. First of all, it allows the independent development and
running of each of the scale-domains. It also allows the incorporation
of additional processes to some particles but ignores them in others.
For the smallest scale, one may desire to include non-Newtonian forces,
such as the Yarkovsky effect \cite{1983ApJ...270..365M} or general
relativity \cite{2014PhRvD..89d4043W,2019RPPh...82a6902D}, whereas for
the largest scale these effects are irrelevant. On the other hand, the
largest-scale may include dynamical friction
\cite{1971ApJ...166..483S} or Milgromian dynamics
\cite{1983ApJ...270..365M,2019MNRAS.484..107S} which are irrelevant
for the microscopic scales.

An addition advantage of the strict separation of scales, it is
possible to operate each system in its own frame of reference and with
a separate coordinate system.  The microscopic and the macroscopic
systems could both operate using a typical lengh-scale and time scale
of order unity. This enables multi-scale simulations that extend
beyond the dynamic range limited by the IEEE double-precision
standard.

\subsection{Second-order non-intrusive coupling}
\label{Sect:ClassicBridge}

In the classic scheme, two different gravitational solvers are
combined to bridge the wide range of scales. The microscopic
scale in \cite{2007PASJ...59.1095F} was a cluster of ${\cal O}(10^3)$
stars in a volume of ${\cal O}(1)$\,pc, whereas the macroscopic scale
was a galactic nucleus of ${\cal O}(10^8)$ stars in a ${\cal
O}(100)$\,pc radius sphere. The relaxation time-scales cover more
than 5 orders of magnitude. In this example, the star cluster was
integrated using a 4-th order Hermite predictor-corrector
direct-summation scheme. The interactions between stars in the galaxy,
and between the cluster and the galactic stars were resolved using a
hierarchical tree-code. Here we briefly discuss the classic coupling
method, as a preparation for the discussion on its high-order
generalization and the augmented bridge (see
\S\,\ref{Sect:Adaptations}).

The \bridge\ integrator can be formulated from a Hamiltonian
splitting argument, in a way similar to the derivation of symplectic
integrators used in planetary dynamics. The Hamiltonian of a
$N$-body system with sub-systems $A$ and $B$ under gravitational
interaction is given by the expression:
\begin{equation}
H = \sum_{i\in A \cup B}^{N} \frac{||\vect{p}_{i}||^{2}}{2 m_{i}} - \sum_{i<j
\in A \cup B}^{N} \frac{G m_{i} m_{j}}{\|\mathbf{r}_{i} -
\mathbf{r}_{j}\|}\,. \label{eq:01}
\end{equation}
Here $G$ is Newton's constant, $m_{i}$, $\vect{r}_i$ and $\vect{p}_i$
are the mass, and vectors for the position and momentum of particle
$i$.

The systems $A$ and $B$ may represent a star cluster and its parent 
Galaxy, respectively. Following \cite{2007PASJ...59.1095F}, the Hamiltonian shown in 
Eq. \ref{eq:01} can be separated in the following way:

\begin{equation}
H= H_{A+B} + H_\textrm{int}=H_A + H_B + H_\textrm{int}, 
\end{equation}
where:
\begin{subequations}
\begin{align}
H_A &= \sum_{i\in A}^{N_A} \frac{||\vect{p}_{i}||^{2}}{2 m_{i}} -
\sum_{i\neq j \in A}^{N_A} \frac{G m_{i} m_{j}}{\|\mathbf{r}_{i} - \mathbf{r}_{j}\|}\,, \\
H_B &= \sum_{i\in B}^{N_B} \frac{||\vect{p}_{i}||^{2}}{2 m_{i}} -
\sum_{i\neq j \in B}^{N_B} \frac{G m_{i} m_{j}}{\|\mathbf{r}_{i} - \mathbf{r}_{j}\|}\,, \\
H_\textrm{int} &= -\sum_{\substack{i\in A, \\
j\in B}}^{N_{A+B}}
\frac{G m_{i} m_{j}}{\|\mathbf{r}_{i} - \mathbf{r}_{j}\|}\,. 
\end{align}
\end{subequations}
In the last equations we do not require to explicitly state $j<i$
because $i$ and $j$ are from separate sets. The time evolution of the
whole system can be written, for a second order approximation, as
follows:
\begin{eqnarray}
\label{eq:bridge}
e^{\tau H} & \simeq & e^{\frac{\tau}{2} H_{A+B}} e^{\tau H_\textrm{int}} e^{\frac{\tau}{2} H_{A+B}} \qquad \text{or} \nonumber\\ 
e^{\tau H} & \simeq & e^{\frac{\tau}{2} H_\textrm{int}} e^{\tau H_{A+B}} e^{\frac{\tau}{2} H_\textrm{int}}\, 
\end{eqnarray}
Here $\tau$ corresponds to the time step in which the two methods are
coupled.

By defining we write the drift and kick operators as:
\begin{subequations}
\begin{align}
D(\tau) &\equiv \prod_{k}^{Q} e^{\tau H_{A+B}}\,\label{eq:op1}
\qquad \text{and} \\ 
K(\tau) &\equiv \prod_{k \neq l}^{Q} e^{\tau H_\textrm{int}}\,,\label{eq:op2}
\end{align}
\end{subequations}
we then rewrite Eq.\,\ref{eq:bridge} as
\begin{eqnarray}
{\cal B}_{2}(\tau) &=& D(\frac{\tau}{2})K(\tau)D(\frac{\tau}{2}),
\qquad \text{or} \nonumber\\ 
{\cal B}_{2}(\tau) &=& K(\frac{\tau}{2})D(\tau)K(\frac{\tau}{2}),
\label{eq:06}
\end{eqnarray}
respectively, where we wrote the bridge operator as ${\cal
B}_{2}(\tau)$, indicating that it is second order and dependent on
the bridge time step $\tau$.

Since $H_\textrm{int}$ depends on the positions only, the operator $e^{\tau
H_\textrm{int}}$ represents a pure momentum kick. During this process, the
velocity of elements in the microscopic system is updated with the
external force generated by the macroscopic system. To prevent
confusion in terms of micro and macro, we tend to refer to the more
astronomical application in which the microscopic system could be
referring to the stars in a cluster whereas the macroscopic system
refers to the cluster as part of the Galaxy. In
fig.\,\ref{fig:ssm_bridge} we presented this view schematically in a
scale separation map. The velocities of the stars in the galaxy are
also updated after computing the acceleration due to their
self-gravity using the tree code.

Since $H_A$ and $H_B$ are completely independent, the evolution
operator $e^{\tau H_{A+B}} \equiv e^{\tau H_{A}} e^{\tau H_{B}}$
consists of the separate evolution of the two subsystems. A full
time-step in \bridge\, then consists of
\begin{itemize}
\item[i)] mutually kicking the sub-systems $A$ and $B$ for 
$\tau/2$,
\item[ii)] evolving the two sub-systems $A$ and $B$ in isolation for
$\tau$ using suitable codes together with an update of their
positions, and
\item[iii)] mutually kicking the 
sub-systems $A$ and $B$ for another $\tau/2$.
\end{itemize}

The strict separation of the operator in Eq.~\ref{eq:bridge} allows us
to solve both parts separately but combine the result to a
self-consistent solution of the whole combined system. This enables
us to split any compound solver to be separated into fundamental parts
that can be solved individually and subsequently combined. This again
allows us to write efficient and confined solvers for each of these
individual parts. Even though depending on the adopted integrator each
of the coupled codes scales as ${\cal O}(N^2)$ or ${\cal
O}(N\log(N))$, the coupling itself scales with ${\cal O}(N)$.

The modularization achieved enables a more efficient calculation of
the evolution of the joined system under the condition that the time
step of the macroscopic system (the interaction term $H_\textrm{int}$)
exceeds that of any of the microscopic time steps (of $H_A$ and $H_B$
systems). This is the case, but not exclusively so, if the spatial
and temporal scales of the microscopic system are well separated from
the macroscopic system. Once this condition is met, it is possible to
integrate the microscopic and the macroscopic systems with different
integrators geared towards their respective requirements. To this
point, the method we described is not different from the classic
bridge method \cite{2007PASJ...59.1095F}.

It is not difficult to find a counterexample where the
\bridge\ integrator degenerates: take e.g. a star cluster where the
stars are assigned to system $A$ and $B$ at random. In this case, the
formal splitting is still valid, but the bridge time step $\tau$
reduces to the global minimum time step. Note that in the approach
above, the coupling strategy is defined manually at the beginning of
the simulation and therefore the coupling remains static throughout
the time evolution of the system. If a merger of the two clusters
occurs during the simulation, the \bridge\ scheme evolves into the
degenerate state. The degeneration of the solver can be mitigated by
allowing the coupling to be time-dependent and variable and if we
allow the two systems to exchange particles at such intervals. A
similar strategy was adopted by \cite{2015ComAC...2....6I} but then
coded directly in C.

\subsection{High-order non-intrusive coupling}
\label{sec:Highorderbridge}

The classic \bridge\ scheme may experience numerical difficulties when
the spatial and/or temporal scales of an interaction of two or more
sub-systems become comparable. This may in part be resolved by
increasing the order of the coupling strategy. Increasing the
bridge-order also allows us to use higher-order dedicated integrators
without losing precision in the bridge step. We present a
generalization of the classic \bridge\, scheme to an arbitrary number
of systems and high orders. We begin by assuming a system of
particles, $S=\bigcup_{k} S_{k}$, composed by a number $Q$ of
sub-systems $S_{k}$.
In this case the total Hamiltonian of the system,
\begin{equation}
H = \sum_{i\in S}^{N} \frac{||\vect{p}_{i}||^{2}}{2 m_{i}} - \sum_{i\neq j \in
S}^{N} \frac{G m_{i} m_{j}}{\|\mathbf{r}_{i} -
\mathbf{r}_{j}\|}\,,\label{eq:04}
\end{equation}
can be split such that we obtain:
\begin{equation}
H= \sum_{k}^{Q} H_{S_{k}} + \sum_{k \neq l}^{Q} H_{S_{k}
S_{l}}^\textrm{int}\,. \label{eq:highO}
\end{equation}
The terms in Eq. \ref{eq:highO} are given by the following relations:

\begin{subequations}
\begin{align}
H_{S_{k}} &= \sum_{i\in S_{k}}^{N} \frac{||\vect{p}_{i}||^{2}}{2
m_{i}} - \sum_{i\neq j \in S_{k}}^{N} \frac{G m_{i}
m_{j}}{\|\mathbf{r}_{i} - \mathbf{r}_{j}\|}\,, \\
H_{S_{k}S_{l}}^\textrm{int} &= \sum_{\substack{i\in S_{k}, \\
j\in S_{l}}}^{N} -\frac{G m_{i}
m_{j}}{\|\mathbf{r}_{i} - \mathbf{r}_{j}\|}\,. \label{eq:05}
\end{align}
\end{subequations}
Based on this splitting, a general, multi sub-system, second-order
time-evolution operator can be constructed, as we demonstrated in
\S\,\ref{Sect:ClassicBridge}.

Similarly to the classic \bridge, operators $e^{\tau H_{S_{k}}}$
independently evolves each of the sub-systems $S_{k}$ in isolation.
Operators $e^{\tau H_{S_{k} S_{l}}^\textrm{int}}$ represents the pure
momentum kicks due to the interaction between sub-systems $S_{k}$ and
$S_{l}$. In the case of the Hamiltonian in eq.~\ref{eq:highO}, the
forces due to the interaction terms do not depend on velocities,
therefore, the operators $e^{\tau H_{S_{k} S_{l}}^\textrm{int}}$ are
commutative. We note, however, that commutability is not possible for
velocity-dependent forces and therefore, a special treatment is
required (see section~\ref {sec:PNBridge} and for an example section
\ref{Sect:dynamicalFriction}). Each of the operators $e^{\tau
H_{S_{k}}}$ and $e^{\tau H_{S_{k} S_{l}}^\textrm{int}}$ can be
associated to different solvers running concurrently.

A high-order \bridge\ scheme can be constructed in a similar way as in
a symplectic integrator. Eq.~\ref{eq:06} can be extended to a higher
order by composition of $D(\tau)$ (see eq.\,\ref{eq:op1}) and
$K(\tau)$ (see eq.\,\ref{eq:op2}) operators \citep{Hairer2005}. For a
$4^\textrm{th}$ symmetric composition with 4 stages, the high-order
\bridge\ takes the form (see also eq.\,\ref{eq:B4}):
\begin{equation} 
\begin{split}
{\cal B}_{4}(\tau) = D(u_0 \tau ) K(u_0 \tau ) D(u_1 \tau) K(v_{1} \tau ) 
D(u_2 \tau ) K(v_1 \tau ) D( u_1 \tau ) K(v_0\tau) D(u_0\tau)\,,\label{eq:09}
\end{split}
\end{equation}
for convenience we list the coefficients for $u_i$ and $v_i$ for a
selection of choices in Tab.\,\ref{Table:Coefficients} of
\S\,\ref{Sect:AppendixA}, but for a more complete overview we refer to
\citep{1990PhLA..150..262Y} and \cite{Hairer2005}. For a sixth order
symmetric composition, $D(\tau/2) K(\tau) D(\tau/2)$ leads to (see
also eq.\,\ref{eq:B6})
\begin{equation}
\begin{split}
{\cal B}_{6}(\tau) = D( w_0/2 \tau ) K(w_0 \tau ) D( (w_0 + w_1)/2 \tau)\ ... \\
...\ D( (w_{s-1}+w_{s})/2 \tau ) K(w_{s} \tau ) D( w_{s}/2 \tau),
\label{eq:10}
\end{split}
\end{equation}
with coefficients $w_i$ given in Tab.\,\ref{Table:Coefficients} (see
also \cite{doi:10.1080/10556780500140664}, their Equations 11 to
17). We present a complete listing of the bridge equations in
eq.\,\ref{eq:B2} to eq.\,\ref{eq:B10}. The self-adjoined methods
associated to eqs.~\ref{eq:09} and \ref{eq:10} are also possible. The
formulation above provides a fully symplectic time evolution if the
codes being bridged are symplectic as well.

Using this scheme, integrators of different orders can be coupled to
construct a high-order scheme by matching the order of the \bridge\ to
be used during the coupling. For example, when coupling a sixth-order
to a fourth-order method, it is probably not appropriate to choose
${\cal B}_{2}(\tau)$, but rather ${\cal B}_{4}(\tau)$ or ${\cal
B}_{6}(\tau)$ to have a convergent compound-method of fourth or
sixth order. If the sixth order bridge ${\cal B}_{6}(\tau)$ would be
chosen, the compound method would still be fourth-order
(constrained by the fourth-order sub-integrator). In such a
hierarchical coupling, while formally the overall order of convergence
of the compound solver is limited by the lowest order, locally
sub-systems being evolved with higher-order are still integrated at
this higher order. This can be advantageous if, for example, the
subsystem dominates in the overall error.

The disadvantage of a higher-order scheme is the requirement for each
of the sub-integrators to be time-reversible. Tree-codes are not
intrinsically time-symmetric and it is hard to make them time
symmetric \cite{2002JCoPh.179...27D}, but for shared time-step higher
order schemes this does not pose a severe limitation (see for example
in \S\,\ref{Sect:HighorderBridge}).

\section{Adaptations to Bridge}\label{Sect:Adaptations}

Sometimes coupling existing methods require additional
non-canonical operations inside the bridge operator. This happens when
the coordinate system changes at run time or if one desires to adopt
specific boundary conditions, which may change at run time.

\subsection{Introducing boundary conditions or an expanding coordinate system}

Changing the coordinate system and invoking periodic boundary
conditions at run-time are common in cosmological simulations to mimic
the expansion of the Universe while limiting the computational domain.
Instead of incorporating these in the physics solver directly, we may
want to implement them in the operator. The same procedure is then
ported from the $N$-body code to the bridge, for example by
introducing periodic boundary conditions in the particle's coordinate
system
\begin{eqnarray}
\dot{\mathbf{r}} & = & \mathbf{v}, \\
\dot{\mathbf{v}} & = & \left(\frac{\ddot a}{a} + {\cal H} \right)\mathbf{r} + \mathbf{g},
\end{eqnarray}
Here ${\cal H}$ could be zero or it could represent the Hubble flow
${\cal H} = 0.5 \frac{H^2_0 \Omega_m}{s^3}$, where the scale factor
$s$, Hubble constant $H_0$ and matter density $\Omega_m$ have their
usual meaning, and $\mathbf{g}$ is Newtonian gravity. A cosmological
bridge can then be constructed by including the terms $\mathbf{r}
\ddot a/a$ in the kick operator.
One argument for adopting \bridge\, to include such boundary
conditions or phase-space alterations is the convenience of using an
unaltered integrator for the drift operator. The performance is not an
issue here because the cosmological terms are not expensive to
calculate and the drift operation scales linearly with the number of
particles in the simulation.

\subsection{\bridge\ in rotating reference frames} 
\label{sec:RotBridge}

Another class of problems require solving on a non-inertial frame of
reference, but, for example, in a rotating frame of reference. We
encountered such a situation when integrating the equations of motion
of a star cluster in orbit around the Galactic center. In that case,
we are interested in the stimulated evaporation of a star cluster due
to the non-radial structure in the Galactic potential. This potential
is implemented as a semi-analytic background with a non-radial structure
in the form of spiral arms and a bar. The bar and spiral arms in the
model rotates as rigid bodies with a particular pattern speed
\citep{2010ApJ...722..112M}. As a consequence, the potential
associated with the various components in the Galaxy model depend on
time. Instead of having the Galaxy rotate, we may opt for
implementing the rotation in the coupling pattern. This results in
better energy conservation for the same time-steps size.
\cite{2016MNRAS.457.1062M}.

We formulate a \bridge\, for a rotating frame of reference, such that
the interactions between the stellar systems and the terms in the
equations of motion arising from the non-inertial terms are bridged
(The latter is convenient because it allows the integral to be
formulated for an inertial frame of reference without adjustments).

The rotating \bridge\ is derived by considering a particle of mass $m$
located in a frame that rotates around the $z$-axis with constant
angular speed $\Omega$. The Hamiltonian of this particle is then
\begin{subequations}
\begin{align}\label{Eq:hamilton1}
H &= \frac{||\vect{p}||^2}{2m} + U_\mathrm{gen}(\vect{r}, \vect{p}), \\
H &= \frac{||\vect{p}||^2}{2m} + U_\mathrm{ext}(\vect{r}) - \left( \Omega \times \vect{r}
\right)\cdot \vect{p}-\frac{1}{2}m ||\Omega \times \vect{r}||^2 .
\label{Eq:hamilton2}
\end{align}
\end{subequations}
Here \vect{r} and \vect{p} are the position and momentum vectors of
the particle in the rotating frame. The term
$U_\mathrm{ext}(\vect{r})$ is the potential energy due to an external
force, which depends only on the position of the particle. For example,
$U_\mathrm{ext}(\vect{r})$ represents the galactic
potential. The last two terms in Eq. \ref{Eq:hamilton2} correspond to
a potential energy which accounts for the centrifugal and Coriolis
forces. The energy associated to the centrifugal and Coriolis forces
together with $U_\mathrm{ext}(\vect{r})$, represents the total
generalized potential energy of the particle,
$U_\mathrm{gen}(\vect{r}, \vect{p})$.

Here $U_\mathrm{gen}(\vect{r}, \vect{p})$ depends on the momentum of
the particle. Therefore, it is not possible to split the above
Hamiltonian to obtain the drift and kick operators as we demonstrated
in Eqs. \ref{eq:op1} and \ref{eq:op2}. We construct a rotating
\bridge\ integrator by splitting the equations of motion of a particle
in such a way that it satisfies Eq. \ref{Eq:hamilton1}.

There are two ways to construct a rotating \bridge. One of them we
call canonical and the other the non-canonical \citep[see
also][]{Pfenniger1993}. In Sects. \ref{sect:canonical} and
\ref{sect:no-canonical} we explain these approaches in more detail.

\subsubsection{Canonical approximation}\label{sect:canonical}

In this approach, the equations of motion of a particle moving in a
rotating frame, are defined in terms of the canonical coordinates
(\vect{Q}, \vect{P}). The canonical momentum (\vect{P}) is defined
as

\begin{equation}\label{Eq:canonicalP}
\vect{P}= \frac{\partial \mathcal{L}}{\partial \vect{\dot{Q}}}=\frac{\partial \mathcal{L}}{\partial \vect{\dot{r}}}= \vect{p} + m(\Omega\times \vect{r}).
\end{equation}
Here $\mathcal{L}$ is the Lagrangian $\mathcal{L}=
\vect{\dot{Q}}\vect{P}- H$. The canonical momentum can be interpreted
as the velocity of the particle seen in the inertial reference frame
which is coaxial to the rotating frame of reference. By using
Eq. \ref{Eq:canonicalP} we can obtain the canonical momenta in
Cartesian components:

\begin{align} 
P_x &= p_\mathrm{x}-m\Omega y, \nonumber \\
P_y &= p_\mathrm{y}+m\Omega x, \nonumber \\
P_z &= p_\mathrm{z}.
\end{align}
The equations of motion of a particle that satisfies Eq. \ref
{Eq:hamilton1} can then be written in terms of the canonical moment as follows:

\begin{align} \label{eq:canonical}
\dot{x} &= p_x/m + \Omega y; \hspace{7mm} \dot{p_x} = F_x +\Omega P_y, \nonumber \\
\dot{y} &= p_y/m - \Omega x; \hspace{7mm} \dot{p_y} = F_y -\Omega P_x, \nonumber \\
\dot{z} &= p_z/m; \hspace{15mm} \dot{p_z} = F_z.
\end{align}
Here $\vect{F}$ is the external force associated to 
$U_\mathrm{ext}(\vect{r})$. We proceed to split Eqs. \ref{eq:canonical} 
to build the kick and drift operators. The set of equations that 
represent the kick operator $K(\tau)$ is
\begin{align} 
\dot{x} &= \dot{y}= \dot{z}=0, \nonumber \\ 
\dot{p_x} &= F_x +\Omega P_y, \nonumber \\
\dot{p_y} &= F_y -\Omega P_x , \nonumber \\
\dot{p_z} &= F_z .
\end{align}
The solution of these equations is :

\begin{subequations} \label{eq:kick_canonical}
\begin{align} 
v_{x} (t+\tau) &= \left[ v_{x{}}(t) -\left(\frac{a_y + \Omega^2y}{\Omega} \right) \right]\cos{(\Omega\tau)} \nonumber \label{kc1} \\
& + \left[ v_{y}(t) +\left(\frac{a_x + \Omega^2x}{\Omega} \right) \right]\sin{(\Omega\tau)} \nonumber \\
& + \frac{a_y+ \Omega^2y}{\Omega}, \\
v_{y} (t+\tau) &= -\left[ v_{x}(t) -\left(\frac{a_y + \Omega^2y}{\Omega} \right) \right]\sin{(\Omega\tau)} \nonumber \\
& + \left[ v_{y}(t) +\left(\frac{a_x + \Omega^2x}{\Omega} \right) \right]\cos{(\Omega\tau)} \nonumber \\
& - \frac{a_x+ \Omega^2x}{\Omega}, \label{kc2} \\
v_{z} (t+\tau) &= v_{z} (t) + a_z\tau. \label{kc3}
\end{align}
\end{subequations}
Here $a_x$, $a_y$ and $a_z$ are the acceleration on a particle in the
various cartesian coordinates, $x$, $y$ and $z$, respectively.

The drift operator $D(\tau)$ in the canonical 
approximation is represented by the following set of equations:

\begin{align} 
\dot{p_x} &= \dot{p_y}= \dot{p_z}=0, \nonumber \\ 
\dot{x} &= p_x/m + \Omega y, \nonumber \\
\dot{y} &= p_y/m - \Omega x, \nonumber \\
\dot{z} &= p_z/m .
\end{align}

The solution of these equations is

\begin{subequations} \label{eq:drift_canonical}
\begin{align} 
x (t+\tau) &= \left[ x(t) -\frac{v_y}{\Omega} \right]\cos{(\Omega\tau)} \nonumber \label{dc1} \\
& + \left[y(t) + \frac{v_x}{\Omega} \right]\sin{(\Omega\tau)} + \frac{v_y}{\Omega}, \\
y (t+\tau) &= -\left[ x(t) -\frac{v_y}{\Omega} \right]\sin{(\Omega\tau)} \nonumber \\
& + \left[ y(t) +\frac{v_x}{\Omega} \right]\cos{(\Omega\tau)} - \frac{v_x}{\Omega}, \label{dc2} \\
z (t+\tau) &= z(t) + v_z\tau. \label{dc3}
\end{align}
\end{subequations}
The canonical formulation has two advantages: it generates a stable
algorithm and this approximation is symplectic (see
Fig. \ref{fig:approx}). However, for systems with interacting
particles, it is convenient to have a drift operator that is
independent of $\Omega$. This is not the case for the canonical
formulation (although this can be remedied by further splitting the
operator).

\subsubsection{Non-canonical approximation} \label{sect:no-canonical}

In the non-canonical approximation, the motion of a particle is defined 
in terms of its position and velocity coordinates (\vect{r}, \vect{v}). 
Given the generalized force $\vect{F}_\mathrm{gen}= 
m\vect{a}-m\Omega\times(\Omega\times\vect{r}) -2m(\Omega\times\vect{v})$,
the equations of motion of a particle in a rotating frame can be 
written as:

\begin{align} \label{eq:Nocanonic}
\dot{x} &= v_x; \hspace{7mm} \dot{v_x} = a_x +\Omega^2x + 2\Omega v_y, \nonumber \\
\dot{y} &= v_y; \hspace{7mm} \dot{v_y} = a_y +\Omega^2y -2\Omega v_x, \nonumber \\
\dot{z} &= v_z; \hspace{7mm} \dot{v_z} = a_z. 
\end{align}

We split Eqs. \ref{eq:Nocanonic} to build the kick and drift operators. 
The set of equations that represent the kick operator $K(\tau)$ is the 
following:

\begin{align}
\dot{x} &= \dot{y}= \dot{z}=0, \nonumber \\
\dot{v_x} &= a_x + \Omega^2x +2\Omega v_y, \nonumber \\
\dot{v_y} &= a_y +\Omega^2y -2\Omega v_x, \nonumber \\
\dot{v_z} &= a_z. 
\end{align}

The solution of these equations give expressions for the kick velocity
of the particle in the three Cartesian coordinates at time $t+\tau$:

\begin{subequations} 
\label{eq:kick_nocanonical}
\begin{align} 
v_{x} (t+\tau) &= \left[ v_x(t) -\left(\frac{a_y + \Omega^2y}{2\Omega} \right) \right]\cos{(2\Omega\tau)} \nonumber \label{v1} \\
& + \left[ v_y(t) +\left(\frac{a_x + \Omega^2x}{2\Omega} \right) \right]\sin{(2\Omega\tau)} \nonumber \\
& + \frac{a_y+ \Omega^2y}{2\Omega}, \\
v_{y} (t+\tau) &= -\left[ v_x(t) -\left(\frac{a_y + \Omega^2y}{2\Omega} \right) \right]\sin{(2\Omega\tau)} \nonumber \\
& + \left[ v_y(t) +\left(\frac{a_x + \Omega^2x}{2\Omega} \right) \right]\cos{(2\Omega\tau)} \nonumber \\
& - \frac{a_x+ \Omega^2x}{2\Omega}, \label{v2} \\
v_{z} (t+\tau) &= v_{z} (t) + a_z\tau. \label{v3}
\end{align}
\end{subequations}
Here the vector $\vect{a}$ corresponds to the acceleration of the
particle due to the external galactic potential
$U_\mathrm{ext}(\vect{r})$.

The drift operator $D(\tau)$ on the other hand, is represented by the 
following set of equations:

\begin{align} 
\dot{v}_x &= \dot{v}_y= \dot{v}_z=0, \nonumber \\
\dot{x} &= v_x, \nonumber \\
\dot{y} &= v_y, \nonumber \\
\dot{z} &= v_z. 
\end{align}

The solution of these equations for the kick velocity of the
particle in the three Cartesian coordinates is simpler than in the
canonical case (Eqs.\,\ref{eq:kick_nocanonical})
\begin{subequations} 
\label{eq:drift_nocanonical}
\begin{align}
x(t+\tau) &= x(t) + v_{x}(t+ \tau/2)\tau,\\
y(t+\tau) &= y(t) + v_{y}(t+ \tau/2)\tau, \\
z(t+\tau) &= z(t) + v_{z}(t+ \tau/2)\tau. 
\end{align}
\end{subequations} 

By using the non-canonical approximation, a second order rotating
\bridge\ can be constructed as follows
\begin{equation}\label{eq:Rbridge}
{\cal B}^{r}_{2}(\tau) = K(\tau/2)\cdot D(\tau)\cdot K(\tau/2), 
\end{equation}
where the operators $K(\tau)$ and $D(\tau)$ are described by
Eqs. \ref{eq:kick_nocanonical} (see Eq.\,\ref{eq:kick_canonical} for
the canonical case) and \ref{eq:drift_nocanonical} (see also
Eq\,\ref{eq:drift_canonical}) respectively. Consequently, every
$\tau/2$ a star receives a velocity kick due to the external galactic
potential and the position of the star is updated every $\tau$.

The rotating \bridge\ can be generalized to a system of self
interacting particles. The Hamiltonian of a stellar
system $A$ which is located in a frame that rotates around the
$z$-axis with constant angular speed $\Omega$ is given by:
\begin{equation}
H= H_\mathrm{A} + H_\mathrm{int},
\end{equation}
where
\begin{align}
H_\mathrm{A} &= \sum_{i\in A}^{N_A} \frac{||\vect{p}_i||^2}{2m_i} -\sum_{i\neq j\in A}^{N_A} \frac{Gm_im_j}{||\vect{r}_{i}- \vect{r}_j||}, \nonumber \\
H_\mathrm{int} &= \sum_{i\in A}^{N_A} \left[U_{\mathrm{ext}}(\vect{r}_i) - (\Omega\times\vect{r}_i)\cdot\vect{p}_i -
\frac{1}{2}m_i||\Omega\times\vect{r}_i ||^2 \right].
\end{align}

The temporal evolution of the system in a second order approximation
is given by Eq. \ref{eq:Rbridge}, which can be written as a
kick-drift-kick operation:
\begin{equation}
e^{H_\mathrm{int}\tau/2} e^{H_\mathrm{A}\tau} e^{H_\mathrm{int}\tau/2}.
\end{equation}
Here the term $e^{H_\mathrm{int}\tau}$ represents the kick operator $K(\tau)$ while the term $e^{H_\mathrm{A}\tau}$ corresponds to the drift operator $D(\tau)$. For a system of self-interacting particles the drift operator is given by
\begin{subequations}
\begin{align}
\vect{x}(t+\tau) = \vect{x}(t) + \vect{v}(t+\tau/2)\tau, \label{eq:nmx}\\
\vect{v}'(t+\tau/2) = \vect{v}(t+\tau/2) + \vect{a}\tau/2. \label{eq:nmv}
\end{align}
\end{subequations}
The evolution of system $A$ during a rotating \bridge\ time step $\tau$
is composed of the following steps:
\begin{itemize}
\item[i)] At $\tau/2$ the system receives a velocity kick due to the external potential of its parent galaxy (Eqs. \ref{eq:kick_nocanonical}). This velocity is referred to as $\vect{v}(t+\tau/2)$.
\item[ii)] The positions of the stars are updated for a time step
$\tau$ (Eq. \ref{eq:nmx}). In addition, the velocities of the stars
are updated once more after evolving system $A$ by means of direct
$N$-body integration (Eq. \ref{eq:nmv}).
\item[iii)] The system receives a velocity kick for another $\tau/2$
(Eq.\, \ref{eq:kick_nocanonical}). This kick is computed by using
the previous velocity $\vect{v}'(t+\tau/2)$. Note that the only
difference between the rotating and classical \bridge\ is in the
kick operator. In particular, the code evolution operator
$e^{H_\mathrm{A}\tau}$ does not have to be changed.
\end{itemize}
The above procedure can also be applied to a more generalized case in
which there are several self-gravitating systems.

The precision of the rotating \bridge\ can be improved by applying the
drift and kick operators in accordance with Eqs. \ref{eq:09} and
\ref{eq:10} as we discussed in \S\,\ref{sec:Highorderbridge}. A higher-order rotating bridge can be constructed similarly. In
\S\,\ref{Sect:rotatingbridge} we present an example of a rotating
bridge.

\begin{figure}[h!]
\center
\includegraphics[width= 8 cm]{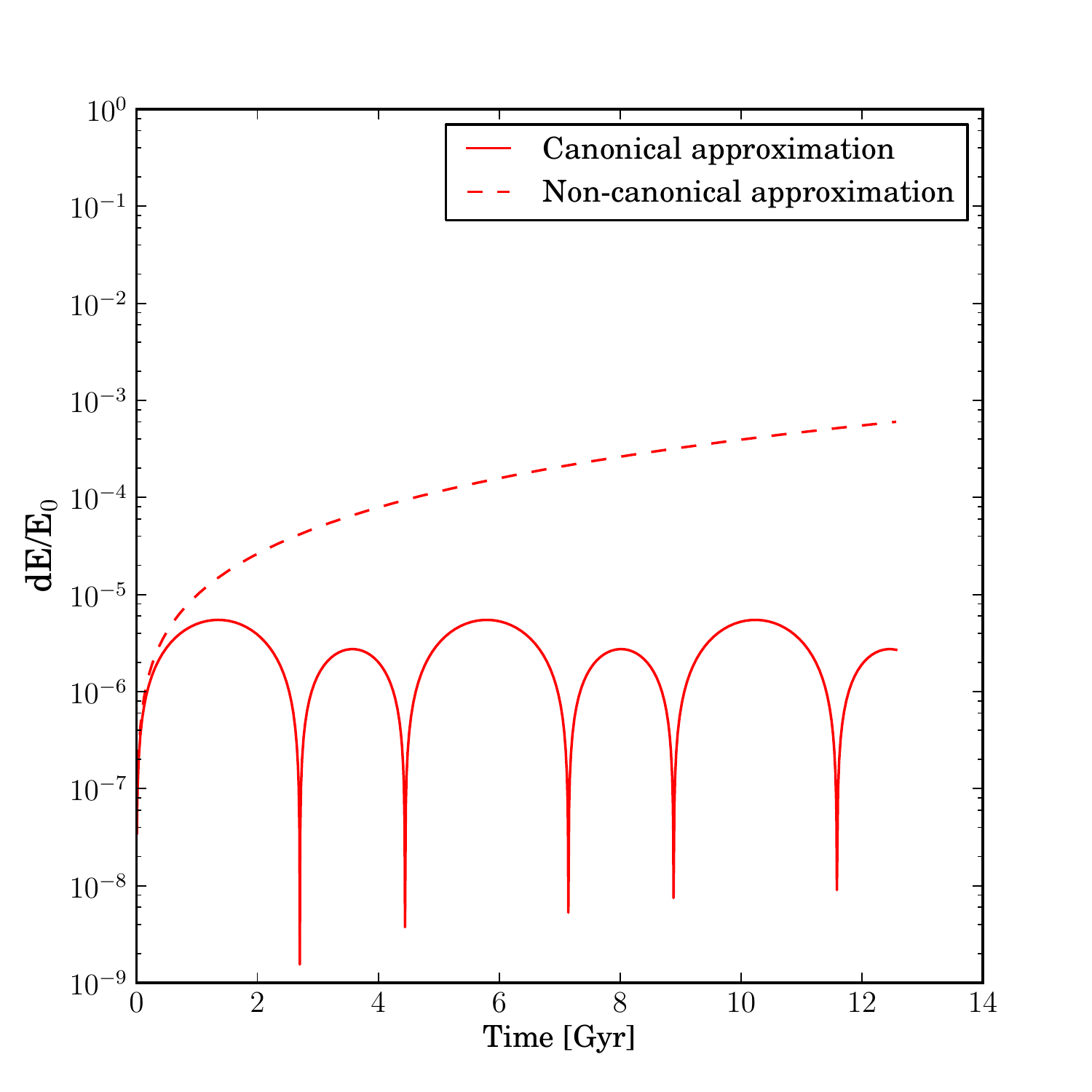}
\caption{Energy error as a function of time for a particle in a
rotating frame. Here we use a second order rotating \bridge\ with
a time step of $1$~Myr.}
\label{fig:approx}
\end{figure}

In fig.\,\ref{fig:approx} we present the energy error resulting from
integrating a test particle in a background potential of a Milky
Way-like galaxy with two spiral arms. One calculation is performed
with the canonical approximation \S\,\ref{sect:canonical} and the
other with the non-canonical approximation
\S\,\ref{sect:no-canonical}. The former (canonical case) is shown to
behave symplectic in terms of energy conservation, whereas the
non-canonical case is not symplectic.

\subsection{\bridge\ expansions for a particular subset}
\label{sec:PNBridge}

In some cases, one would like to apply a certain force or operation to
a specific subset of particles. For example, in the
Yarkovsky-O'Keefe-Radzievskii-Paddack \citep[YORP for
short,][]{2000Icar..148....2R} effect and the diurnal Yarkovsky
effect \citep{2015aste.book..509V} only the lightest pebbles are
affected, whereas planets and other major bodies are not. One could
solve this by integrating pebbles with a separate integrator in which
a direct $N$-body code is bridged with the YORP calculation and
another $N$-body code for the planets. This, however, makes it harder
to deal with inter-code operations, such as close encounters between
planets and asteroids, tidal effects of collisions. With two separate
$N$-body codes, detecting a collision can then be implemented in the
bridge step, but this is generally longer than the internal
integration time scale of any of the bridged sub-codes. This
limitation can be overcome by reducing the bridge time step $\tau$,
but only at the cost of spending more time in the bridge, which is
generally expensive compared to the optimized $N$-body codes (See also
the discussion in chapter 4 of \cite{AMUSE}). A better solution is to
augment a subset of the bridged particles with the additional physics,
whereas the other subset is not affected.

Here we demonstrate how to construct such an integrator.
We provide the example in which a small
subset of particles is affected by post-Newtonian dynamics, whereas
the majority of the particles is integrated under the Newtonian
approximation.

In order to extend \bridge\ to include post-Newtonian (PN) corrections
a special treatment is needed to handle the velocity dependency in the
PN terms. Here we adopt the recipes developed in
\cite{2010CeMDA.106..143H}, where an auxiliary velocity,
$\mathbf{w}_{i}(t)$ with $\mathbf{w}_{i}(t=0) = \mathbf{v}_{i}(t=0)$,
is introduced to make the time evolution operators separable and
therefore allowing the use of the leapfrog algorithm for implementing
the PN corrections.

In \bridge, the operators $e^{\tau H_{S_{k}}}$ and $e^{\tau H_{S_{k}
S_{l}}^\textrm{int}}$ in eqs.~\ref{eq:op1} and~\ref{eq:op2} can be
associated with different solvers for each of the sub-systems in a
simulation: For example, if the $j$-th sub-system requires PN
corrections whereas interactions within the other sub-system can be
treated classically. In that case, only the operator $e^{\tau
H_{S_{k}}}\vert_{k=j}$ in eq.~\ref{eq:op1} requires modifications,
in the form of a simple substitution of a PN solver to the regular
Newtonian solver for this particular sub-system $j$.

In the more complex case where several interacting sub-systems require
PN corrections, both operators $D(\tau)$ and $K(\tau)$ in
eqs.~\ref{eq:op1} and~\ref{eq:op2} have to be modified accordingly. In
the following, we will assume that from a total number $Q$ of
sub-systems being bridged, a number $Q^{N}$ are ``Newtonian'',
and a number $Q^{E}$ require PN corrections so that $Q = Q^{N} +
Q^{E}$. We define $S^{N}$ as the set of ``Newtonian''
sub-systems and $S^{E}$ as the set of ``post-Newtonian'' sub-systems,
so that $S = S^{N} \bigcup S^{E}$. In this way, operator $D(\tau)$ in
eq.~\ref{eq:op1} can be extended into the following expression

\begin{equation}
\tilde{D}(\tau) = \prod_{k\in S^{N}}^{Q^{N}} e^{\tau H_{S_{k}^{N}}} \prod_{k\in S^{E}}^{Q^{E}} e^{\tau H_{S_{k}^{E}}}. \label{eq:11}
\end{equation}
Here $e^{\tau H_{S_{k}^{N}}}$ represents a Newtonian solver for the
$k$-th sub-system in $S^{N}$ and $e^{\tau H_{S_{k}^{E}}}$ represents
a PN solver for the $k$-th sub-system in $S^{E}$. With a PN solver we
indicate a code that evolves particles under the total acceleration
$\mathbf{a} = \mathbf{a}^{N} + \mathbf{a}^{E}$, rather than just
$\mathbf{a}^{E}$ as might be implied by our notation.

The operator $K(\tau)$ in eq.~\ref{eq:op2} can be extended as follows. We first define some auxiliary kick operators,

\begin{equation}
K^{N\leftrightarrow N}(\tau) = \prod_{k \neq l}^{Q^{N}} e^{\tau H_{S_{k}^{N} S_{l}^{N}}^\textrm{int}}\,,\label{eq:12}
\end{equation}
\begin{equation}
K^{N\leftrightarrow E}(\tau) = \prod_{k\in S^{N}}^{Q^{N}}\prod_{l\in S^{E}}^{Q^{E}} e^{\tau H_{S_{k}^{N} S_{l}^{E}}^\textrm{int}}\,,\label{eq:13}
\end{equation}
\begin{equation}
K^{E\leftrightarrow E}(\tau) = \prod_{k \neq l}^{Q^{E}} e^{\tau H_{S_{k}^{E} S_{l}^{E}}^\textrm{int}}\,,\label{eq:14}
\end{equation}
from which follows that the extended kick operator for \bridge\ with
PN corrections can be written as (see also eq.\,\ref{eq:B2}):
\begin{center}
\begin{eqnarray}
\tilde{K}(\tau)
&=& K^{N\leftrightarrow N}(\tau/2)\cdot K^{N\leftrightarrow E}(\tau/2)\cdot K^{E\leftrightarrow E}(\tau) \cdot K^{E\leftrightarrow N}(\tau/2)\cdot K^{N\leftrightarrow N}(\tau/2) \nonumber\\
&=& K^{E\leftrightarrow E}(\tau/2)\cdot K^{E\leftrightarrow N}(\tau/2)\cdot K^{N\leftrightarrow N}(\tau) \cdot K^{N\leftrightarrow E}(\tau/2)\cdot K^{E\leftrightarrow E}(\tau/2)\,.
\label{eq:15}
\end{eqnarray}
\end{center}

%\begin{align}
%\begin{split}
%\tilde{K}(\tau) &= K^{N\leftrightarrow N}(\tau/2)\cdot K^{N\leftrightarrow PN}(\tau/2)\cdot K^{PN\leftrightarrow PN}(\tau)\cdot \nonumber\\
% \cdot K^{PN\leftrightarrow N}(\tau/2)\cdot K^{N\leftrightarrow N}(\tau/2) \nonumber\\
%\end{split}
%\begin{split}
%\tilde{K}(\tau) &= K^{PN\leftrightarrow PN}(\tau/2)\cdot K^{PN\leftrightarrow N}(\tau/2)\cdot K^{N\leftrightarrow N}(\tau)\cdot \\
% \cdot K^{N\leftrightarrow PN}(\tau/2)\cdot K^{PN\leftrightarrow PN}(\tau/2)\,.\label{eq:15}
% \end{split}
%\end{align}
%

Eq.~\ref{eq:12} represents the Newtonian kick due to the interaction
between ``Newtonian'' sub-systems, and is identical to the original
definition in eq.~\ref{eq:op2}. Eq.~\ref{eq:13} represents the kick
due to the interaction between ``Newtonian'' and ``post-Newtonian''
sub-systems. In this particular case, a choice has to be made on
whether or not PN corrections should be included. Such a decision
could be based, for example on the distance between the two
interacting subsystems or their masses. Eq.~\ref{eq:14} represents
the PN kick due to the interaction between ``post-Newtonian''
sub-systems. Finally, eq.~\ref{eq:15} represents the extended kick
operator to be used in \bridge\ with PN corrections.

\section{Extension of the construction strategies of \bridge\label{sec:notation}}

\subsection{Extending notation and terminology: the hierarchical coupling
language\label{sec:couplingnotation}}

To accommodate an abstraction to our understanding and
discussion on bridge topologies, we introduce a meta description
language of the coupling patterns using \bridge. We, therefore,
introduce a notation which helps us to think in more abstract terms
about the various coupling strategies. For each of the possible
adaptations, we suggest a notation, examples are given in
\S\,\ref{Sect:classicbridge} to \S\,\ref{Sect:hierarchical}.

In the classic implementation, Fujii et al.\,
\cite{2007PASJ...59.1095F} constructed a bridge between a direct
$N$-body code and a tree-code. In our nomenclature, we called these
codes $S$ for the microscopic system and $G$ for the macroscopic
system. We indicate the order of the integrator using an integer
subscript, $2$ for $2^{\textrm{nd}}$ order, $4$ for $4^{\textrm{th}}$
order, etc. The bi-directional coupling strategy using a
$2^{\textrm{nd}}$ order scheme for the coupling, discussed in
\S\,\ref{Sect:ClassicBridge}, is then written as
\begin{equation}
[ S_4 \lrBridge{2} G_2 ].
\end{equation}
A second-order bridge is appropriate here because the lowest order in
the hierarchy is also second order. Using a $4^\textrm{th}$ bridge
puts specific constraints on the choice of integrators, but will lead
to better energy conservation. In future notation, we tend to omit the
order of the bridge when it is the same as the lowest of the two
coupled codes. An example of the classic bridge is given in
\S\,\ref{Sect:classicbridge}.

Higher accuracy would be acquired in the previous calculations by
adopting an individual time-step direct $N$-body scheme for the global
system. This would motivate the use of a fourth-order bridge
\begin{equation}
[ S_4 \lrBridge{} G_4 ],
\end{equation}
or even $6^{\textrm{th}}$ order for the subsystem
\begin{equation}
[ S_6 \lrBridge{} G_4 ].
\end{equation}
Note that here both bridges are implicitely assumed to be of
$4^{\textrm{th}}$ order. In \S\,\ref{Sect:HighorderBridge} we present
an example of the higher-order bridge coupling.

One could replace the numerical integration by semi-analytic a
potential.
\begin{equation}
[ S_4 \lBridge{2} G ].
\end{equation}
Here the global system has no order because it uses a semi-analytic
potential on which particles float, but they are integrated using the
$2^{\textrm{nd}}$-order bridge. We only presented the bridge with an
arrow in one direction, indicating the single direction of the
hierarchical coupling: the subsystem is affected by the global system,
but not vice versa. The solution is subsequently not self-consistent,
but the calculation will be fast.

\subsection{Augmented coupling strategies}

The effect of the global system can still be taken into account by
including a correction term to the bridge. This additional term comes
in the form of a function $f$, which depends on the characteristics of
the subsystem and the global system. We write the one-directional
bridge with a one-directional support function as
\begin{equation}
[ S \lBridge{f} G]
\end{equation}
Here the function $f \equiv f(S,G)$ below the left-pointed arrow
represents a function that operates on the subsystem (S) with
additional information about the global system (G). It could, for
example, calculate the effect of dynamical friction on the cluster $S$
due to its evolution in the Galaxy $G$. In that case, the function $f$
would contain the dynamical term exerted on the orbit of a star
cluster due to its interaction with the Galactic field stars. The
function $f(S,G)$ is rather general and can include terms for
post-Newtonian correction, the Yarkovsky effect or gas drag. We
present an example of the mono-directional bridge coupling in
\S\,\ref{Sect:dynamicalFriction}.

A special case of an augmented bridge is the rotating bridge. In that
case, the function $f$ describes the consequences of a rotating frame
of reference, as we discussed in \S\,\ref{sec:RotBridge}. An example
is provided in \S\,\ref{Sect:rotatingbridge}.

\subsection{Compound hierarchical coupling strategies}\label{Sect:CompoundHierarchicalBridge}

Instead of adding an interacting function, a new bridge could be
declared to generate a compounded bridge. We could add another
sub-system to the already present system. The potential advantage of a
cascade of subsystems is illustrated by studying the dynamics of
planetary systems of stars in a cluster that is part of a galaxy.
Maybe there are even multiple clusters that interact as part of the
same galaxy. For clarity, we indicate the planetary integrator with
(P, for picoscopic system), the star cluster with (S) and the galaxy
with (G). One way to address such a hierarchical coupling strategy
can be written as
\begin{equation}
[[P_6 \lrBridge{} S_4] \lrBridge{} G_2 ]].
\end{equation}
In this example, the planets (P) and stars (S) integrators are coupled
with a $4^\textrm{th}$-order scheme, whereas the stars in the cluster couple to
the galaxy using a $2^\textrm{nd}$-order scheme. To allow the Galactic
potential to affect the star cluster but not vice versa one could opt
for the more efficient, but inconsistent coupling
\begin{equation}
[[P_6 \lrBridge{} S_4] \gets G_2 ]].
\end{equation}

This scheme can be further expanded hierarchically, for example by
adding a lunar system (M, for minimicroscopic) to one of the planets.
\begin{equation}
[M \gets [P \xleftrightarrow{\hspace*{3mm}}{} [ S \gets G ]]].
\end{equation}
In this case, the moons (M) are affected by the planets (P), the stars
(S) and the Galaxy (G), but the Galaxy is unaware of the moons,
planets or stars. Depending on the requirements for accuracy and speed,
there are several ways in which such a hierarchical compound system
can be constructed, the adopted topology depends on the underlying
scientific question. The topology could be democratic
\begin{equation}
[[M \gets P] \xleftrightarrow{\hspace*{3mm}}{} [S \gets G]],
\end{equation}
or have a reversed hierarchy
\begin{equation}
[[[M \gets P] \xleftrightarrow{\hspace*{3mm}}{} S] \gets G ].
\end{equation}
The various hierarchies and direction of the coupling makes the
technique versatile and flexible. In particular if the implementation
allows changes to the topology at run-time.

\subsection{Bridging of a specific subset}

The notation can be further expanded for compound systems in which one
part ($S$) is affected in a different way than a subset ($S^{\prime}$),
\begin{equation}
[S \gets G, S^{\prime} \lBridge{f} G].
\end{equation}
In this case, $S \subset S^{\prime}$, but this is not necessarily
relevant, in which case one could write:
\begin{equation}
[(S-S^{\prime}) \gets G, S^{\prime} \lBridge{f} G].
\end{equation}

In \S\,\ref{sec:PNBridge} we presented the technical
implementation and illustrated it for the use for including
post-Newtonian corrections in a purely Newtonian N-body solver.

The possibility of declaring specific subsets, combined with
hierarchical couplings, higher-order, and augmentations make for a
powerful non-intrusive tool in which complicated simulation
environments can be constructed. In \S\,\ref{Sect:hierarchical} we
present an example of such a compound hierarchical implementation for
simulating planetary systems in star clusters.

\section{Implementation, validation, verification and Demonstration}
\label{sec:examples}

\subsection{Implementation in the Astrophysical Multipurpose Software Environment}

The wide variety of \bridge\ coupling strategies can be implemented in
several ways. Rigid implementations have been introduced to
simulate planetary systems \cite{2017PASJ...69...81I}, star clusters
\cite{2019ApJ...875...20H} and galactic nuclei
\cite{2015ComAC...2....6I}. We illustrate our implementation as we
have realized in the Astrophysical Multipurpose Software Environment
(\amuse) \citep{2013CoPhC.183..456P,2013AA...557A..84P,AMUSE} and in
\omuse\ \cite{gmd-2016-178}.

\amuse\ is a software environment for astrophysical simulations
written in multiple languages but the fundamental structure is based
on Python \cite[see][]{7328651}. \amuse\ presents a wide variety of
astrophysical codes using homogeneous interfaces, simplifying their
use. \omuse\, is the equivalent of \amuse\, but then for simulating
seas and oceans on the surface of a celestial body.

The \bridge\ method has turned into an essential part for the proper
operation of \amuse\ and \omuse\ which enables us to couple with other
codes from different domains. \amuse\, includes a number of
hydrodynamics and gravitational dynamics codes, for \omuse\ the
coupling is similar.

For the \bridge\ integrators here, it is important to note that the
gravitational dynamics codes in \amuse\ provide convenient
implementations of the evolution operator $e^{\tau H}$ in the form of
an {\tt evolve\textunderscore model} method on the interface. This can
be combined with simple force evaluations (by using another component
code or implemented on the interface level) to provide $e^{\tau
H_\textrm{int}}$, the interaction operators. The integrators presented
above can be quickly formulated using ready-made 'building blocks,'
much in the Duplo philosophy advocated in
\cite{2018Sci...361..979P}. It supports the selection of the
appropriate integrator from a wide variety of implementations to
balance the needs for the precision, accuracy, and performance or by
focusing on some specific characteristics of a particular
implementation.

In the following sections we present several examples of bridged
schemes as they may appear in astrophysics. We complement these tests
with validation.

\subsection{Example 1: the classic bridge: $[S_4 \lrBridge{} G_2]$}\label{Sect:classicbridge}

\fboxsep=8pt\relax
\fboxrule=2pt\relax
\begin{codesnippet}[ht]
%%\begin{figure}
\centering
\begin{boxedminipage}{7.5cm}
{
\footnotesize
\begin{verbatim}
(1) code1=Hermite()
(2) code1.particles.add_particles(cluster)
(3) code2=BHTree()
(4) code2.particles.add_particles(galaxy)
(5) sys=Bridge(timestep=0.1 | units.Myr)
(6) sys.add_system(code1, (code2,))
(7) sys.add_system(code2, (code1,))
(8) sys.evolve_model(10 | units.Myr)
\end{verbatim}
}
\end{boxedminipage}
\caption{ Example usage for a $[S_4 \lrBridge{} G_2]$ bridge using the
{\amuse} framework. }
\label{fig:Bridgeimplementation1}
%\end{figure}
\end{codesnippet}

To illustrate the bi-directional bridge, in
snippet\,\ref{fig:Bridgeimplementation1} we show the usage of the
\bridge\ integrator through the \amuse\ framework. In this example, we
present the steps needed to evolve a star cluster (contained in the
\textrm{cluster} particle set\footnote{A ``Particle set'' is \amuse\,
nomenclature, indicating the collection of individual objects that are
subject to the same force law.}) in its parent galaxy (the
\textrm{galaxy} set). The initial realization can be constructed
within AMUSE using e.g. a Plummer sphere \cite{1911MNRAS..71..460P}
model. In line number (1) we initialize the $N$-body integrator to
calculate the internal evolution of the star cluster. In this case, we
use the Hermite integrator. In line (2) We send the particle data to
the $N$-body code. (3) and (4) similarly a code appropriate for the
galaxy model (in this case a tree-code) is started and
initialized. (5) We instantiate the \bridge\ integrator, setting a
time step for the coupling timescale. We couple the cluster code and
the galaxy into \bridge\ in lines (6) and (7). The method {\tt add
\textunderscore system} has two arguments: the main $N$-body
realization (called the \textit{system}) and a set with
\textit{interaction partners}. The \textit{interaction partners}
indicate which systems will kick the \textit{system}. Therefore, in
line (6) the galaxy will kick the particles in the cluster code. In
line (7) the particles in the cluster code will kick the galaxy. In
this way, we ensure that both cluster and galaxy are evolved
self-consistently. Finally, in line (8) the compound system is evolved
for a particular time frame.

\begin{figure}[h!]
\center
\includegraphics[width=0.7\textwidth]{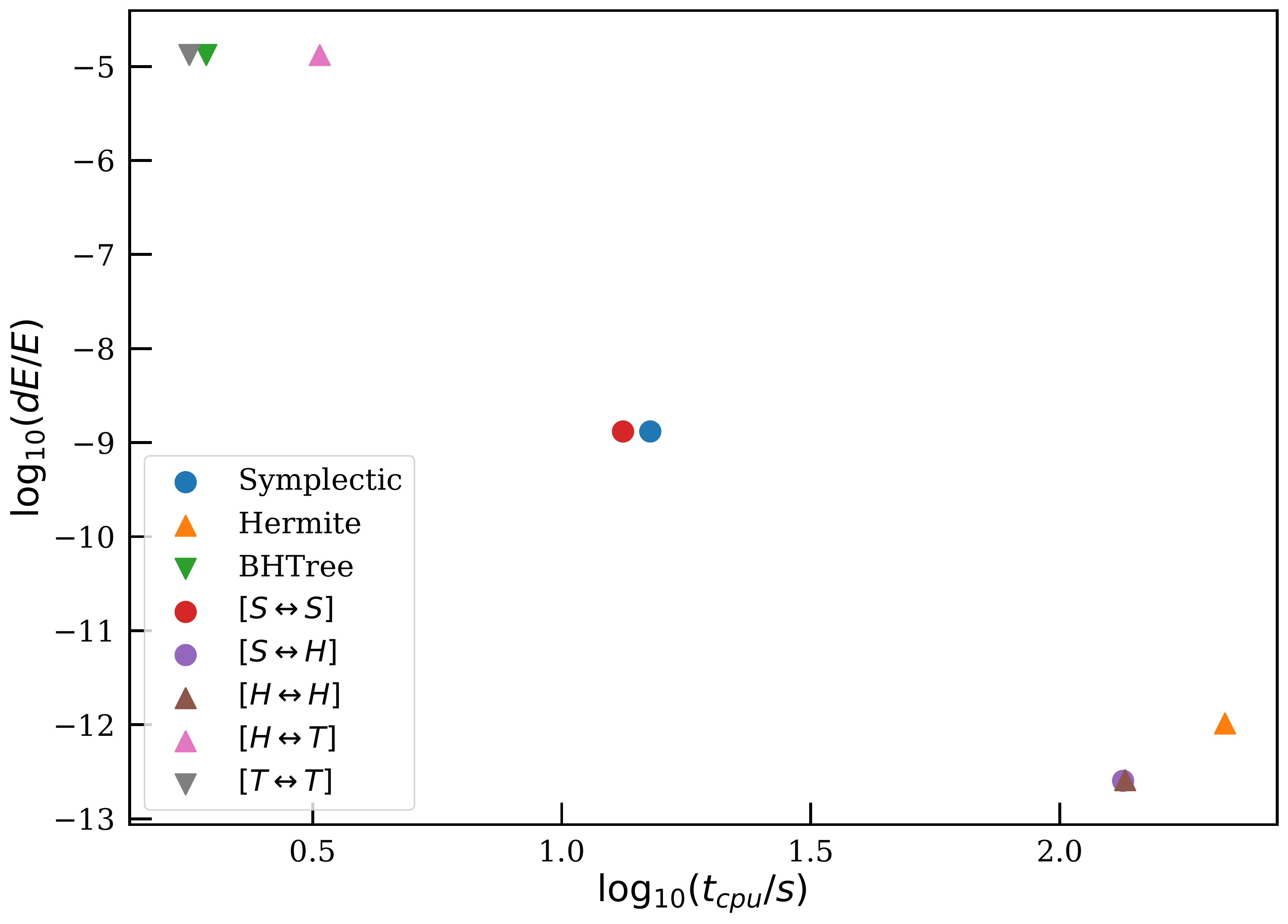}
\caption{Performance in terms of wall-clock time (x-axis) and energy
  conservation (y-axis) for a selection of bridges and codes. In these
  cases, it is not the bridge that gives rise to the energy error, but
  the codes that are coupled. The codes we coupled included the
  $2^\textrm{nd}$-order symplectic integrator {\tt Huayno} (indicated
  as Symplectic and the letter S in the legend), the
  $4^\textrm{th}$-order Hermite predictor-corrector integrator
  (indicated with Hermite and H) and the Barnes-Hut tree code (BHTree
  and T).  The bridged schemes adopt two integrators in a
  bi-directional coupling (see \S\,\ref{sec:couplingnotation}). The
  lowest order integrator determines the overall energy error.  The
  initial conditions for the calculations ware a star cluster of 1000
  stars and a total mass of 600\,\MSun\, in a 10\,pc Plummer sphere
  orbiting in a $10^{11}$\,\MSun\, life Galaxy composed of $10^4$
  equal-mass stars distributed in a standard bulge-disk-halo structure
  with a characteristic radius of 10\,kpc generated using the {\tt
    Galaxia} \citep{2011ascl.soft01007S} software in AMUSE.  The
  globular cluster was put in a circular orbit at a distance of 5\,kpc
  from the Galactic center. We adopted a bridge time-step of 0.1\,Myr
  and integration lasted for 100\,Myr for each of the simulations.}
\label{fig:bridge_comparison_performance}
\end{figure}

The coupling strategy provides a symplectic time evolution. We
demonstrate this in fig.\,\ref{fig:bridge_comparison_performance},
were we measure the energy conservation of several integrations using
\bridge\ .  This example is for illustrative purposes, but it is
apparent that in all cases the bridged methods requires less computer
time at the same energy conservation. When the underlying bridged
systems become progressively more complicated and expensive in terms
of computer time, the relative speed increase of the bridge improves
whereas the energy conserving characteristics remains roughly the
same. This is just a small non-optimal example of the working of a
bridge. In this case, it indicates that the speed does not necessary
affect the energy conservation. The relative efficiency in terms of
performance and accuracy depends in various factors such as the
topology of the system, the selected codes and the tuning parameters
in each of the individual codes. It goes too far for this paper to
explore those parameters, as they should be tuned for each individual
problem separately. Regretfully, we have to ready solution yet for the
optimal choices of these parameters.

The various measurements presented in
fig.\,\ref{fig:bridge_comparison_performance} indicate that the error
in the energy is dominated by the lowest-order methods adopted in the
bridge. The energy errors introduced in the high-order integrator of
the microscopic system is negligible compared to the error introduced
by the low-order integrator used for the macroscopic system. Also the
error in the energy introduce by the bridge method itself is
negligible. The choice of parameters does not emphasize the speedup
introduce by the splitting method, simply because most of the work in
these cases is done in the macroscopic system. Further fine-tuning
will allow the user to make a trade-off between speed and accuracy,
depending on the specific requirements of the problem.

The formulation is not limited by two sub-systems, but can be composed
of an arbitrary number of subsystems. Multiple sub-systems can be
integrated using different specialized solvers and bridged either
using a Hamiltonian splitting technique (as we discuss in section
\ref{sec:Highorderbridge}) or by applying the above splitting scheme
recursively \citep[this was done in ][]{2012MNRAS.420.1503P}. Besides,
multiple microscopic systems may be nested hierarchically, in which
case the macroscopic system can interact with all or only with a
subset of the microscopic systems.

\subsection{Example 2: higher order bridge: $[S_6 \lrBridge{} G_4]$, $[S_8 \lrBridge{} G_6]$, etc.}
\label{Sect:HighorderBridge}

In this example, we calculate the evolution of a stable hierarchical
quadruple system consisting of two binary stars that orbit each
other. The total system comprises of 4 equal-mass bodies \citep[for a
total mass of 1 in $N$-body units,][]{1986LNP...267..233H}, and the
orbits are co-planar with the two binary orbits having a semi-major
axis $a=1/8$ and moderate eccentricity ($\epsilon=0.5$). Each binary
is set-up in an orbit with $a=1.$ and $\epsilon=0.5$. Each binary is
integrated with separate codes and their interaction is bridged using
the $2^\textrm{nd}$-order \bridge\,, a fourth-order, sixth-order, and a tenth-order
\bridge. The binaries themselves are evolved using a Kepler solver,
which calculates the evolution of the subsystems to machine precision.
As a consequence, the measured error is caused by the \bridge\, at
least down to a relative energy error of $dE/E \sim 10^{-15}$.

\begin{figure}[h!]
a)\includegraphics[width=0.5\columnwidth]{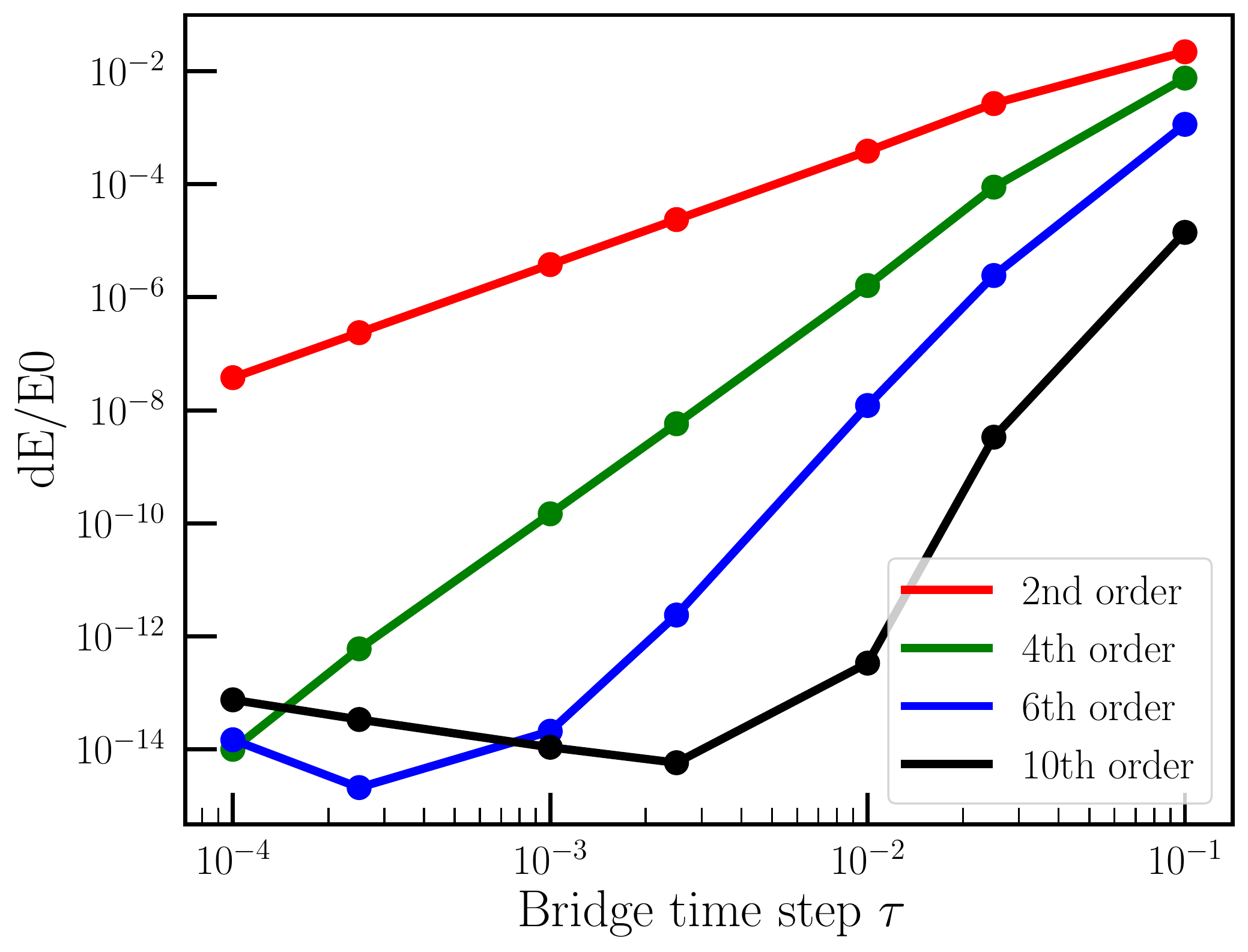}
b)~\includegraphics[width=0.5\columnwidth]{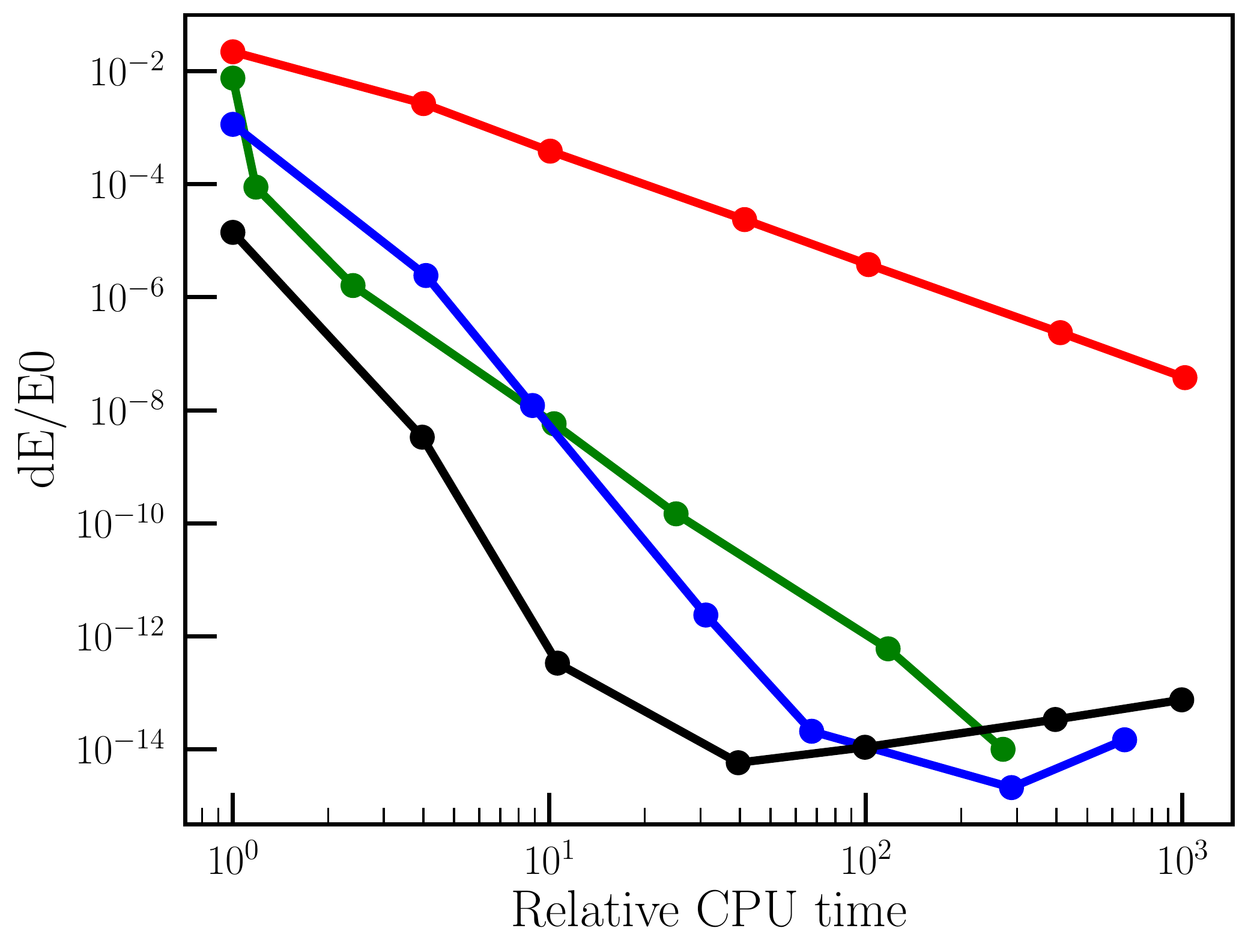}\\
\caption{a) Energy error for integrating two binaries in orbit around
each other using a bridge. The lines give the relative energy error
of the integration as a function of the bridge time step $\tau$ in
dimension-less $N$-body units. From top to bottom, the lines red,
green, blue and black give the relative energy error for a second,
fourth, sixth and tenth order \bridge\, coupling (see also the
legend in Fig.\,\ref{fig:kingtest}).
b) CPU time for the same calculations as in the other panel. The
CPU time is normalized to the lowest accuracy calculations, because
for the adopted size of the system the overhead of the bridging
method is large compared to the actual time spent in the
integrators.
To prevent clutter, we present the legend in the inset in the
left-hand panel of Fig.\,\ref{fig:kingtest}. }
\label{fig:high_order}
\end{figure}

In figure~\ref {fig:high_order}, we present the relative energy error
as a function of the bridge time step $\tau$ for a series of identical
experiments. The setup is a quadruple star system composed of two
binaries that orbit each other. The two binaries are integrated using
a Kepler solver that, apart from round off, does not produce any
errors. The \bridge\ schemes give the expected error behavior,
appropriate for their respective order (Note that the sixth and tenth
order integrators show saturation around the machine precision, as
expected. Improved precision beyond the common $\sim 16$ decimal
places and $\sim 10^{-12}$ energy conservation per step is hard and
requires special treatment of the force evaluation and time-stepping,
as is advocated in \cite{2015ComAC...2....2B}.). We performed the same
series of experiments using a leapfrog \cite{PhysRev.159.98}
integrator (not show in figure~\ref{fig:high_order}), which at small
time steps ($\aplt 10^{-3}$) produces energy errors $\sim 3$ orders of
magnitude higher than the Kepler solver: the behavior of the \bridge\,
is washed out by the errors produced in the leapfrog integration of
the subsystems. This test demonstrates that the \bridge\ scheme can
be applied effectively to couple codes at higher-order.

\begin{figure}[h!]
%\includegraphics[width=0.5\columnwidth]{./plot_kingtest_top.pdf}
%~\includegraphics[width=0.5\columnwidth]{./plot_kingtest_bottom.pdf}\\
a) \includegraphics[width=0.5\columnwidth]{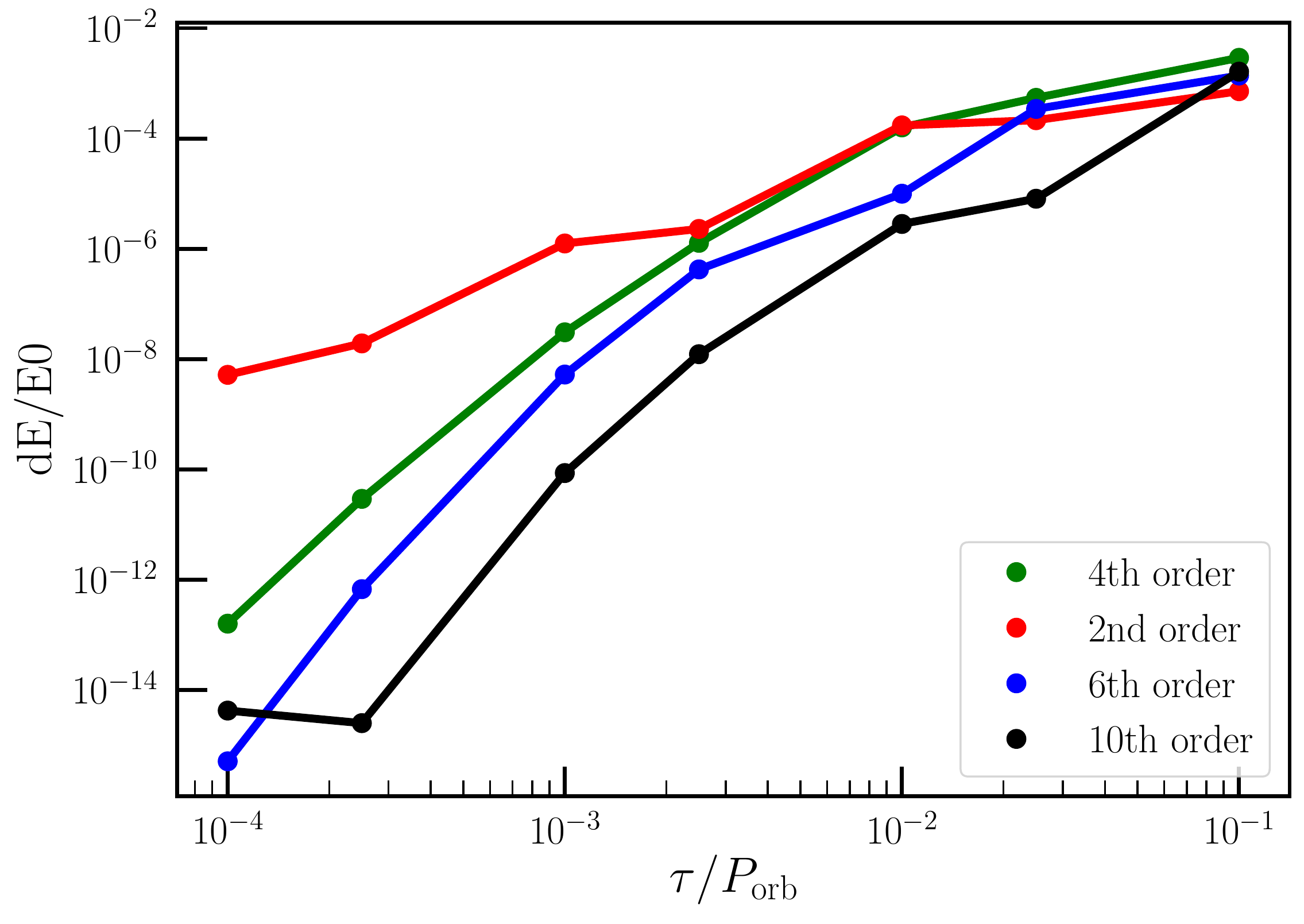}
b) ~\includegraphics[width=0.5\columnwidth]{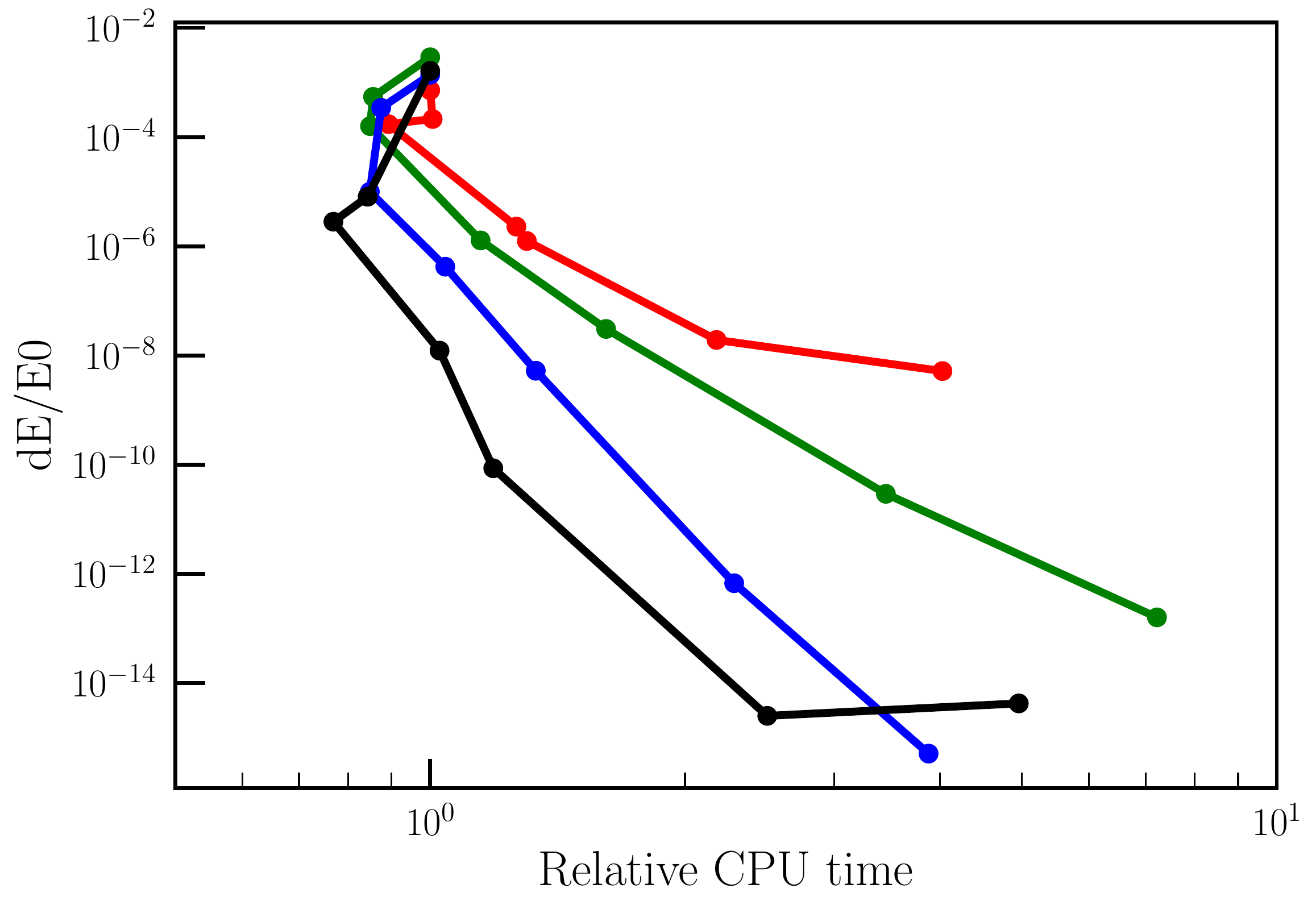}
\caption{a) Energy error for two orbiting clusters clusters of 64
particles. The two clusters are initialized with a King model
\cite{1966AJ.....71...64K}. Plotted is the energy error as a
function of the \bridge\, time step $\tau$ (given as a fraction of
the orbital period of the two clusters). The order of the \bridge\,
is given by the colors, red, green, blue and black for ${\cal
B}_{2}(\tau)$, ${\cal B}_{4}(\tau)$, ${\cal B}_{6}(\tau)$ and
${\cal B}_{10}(\tau)$ order \bridge\, operators.
b) Energy error as a function of the relative CPU times for the same
calculations. As in Fig.\,\ref{fig:high_order} (panel b), the CPU
time is normalized to the lowest accuracy calculations, because for
the adopted size of the system the overhead of the bridging method
is large compared to the actual time spent in the integrators.}
\label{fig:kingtest}
\end{figure}

In figure~\ref{fig:kingtest} we show the energy error for a similar
test where we put two star clusters in orbit around each other. The
models consist of King models with 64 particles, put on circular
orbits with a separation of 8 $N$-body scale-length units. To speed-up
the calculation, we introduce a softening length of 0.2 $N$-body
scale-length units \cite{2012MNRAS.425.1104B}. We integrate for half
an orbital period (which is 284 $N$-body time units). The two models
are integrated using a high precision sixth order method which are
bridged using either ${\cal B}_{2}(\tau)$ (red), ${\cal B}_{4}(\tau)$
(green), ${\cal B}_{6}(\tau)$ (blue) and ${\cal B}_{10}(\tau)$
(black). For the $10^{\textrm{th}}$-order integrators it is hard to
establish the order from the figure because the energy error decreases
so steeply that we quickly run in the round-off error of the
computer. Overall, however, the energy error shows the expected
behavior figure~\ref{fig:kingtest}.  These high-order methods would
benefit from using extended or arbitrary-precision arithmetic, such as
is realized in the calclations in \cite{2015ComAC...2....2B}. It
appears, however, that for most applications (in astrophysics) double
precision-arithmetic is sufficient \cite{2018CNSNS..61..160P}.

Eventually, it is up to the reader to decide if spending more computer
time is worth the effort on reducing the energy error. The right-hand
panels in Figs.\,\ref{fig:high_order} and \ref{fig:kingtest} we
present the relative time spend in the various integrators. For the
few particles in the examples, the higher-order integrators turn out to
be considerably more expensive than the lower-order methods. When the
number of particles (and therefore the number of operations) both
sub-systems increases the relative overhead of adopting a higher-order
integrator drops. In practice, it is probably the lowest order in the
individual integrators that determines the choice for the order of the
bridge integrator, rather than the computational cost.

\subsection{Example 3: augmented bridge to include a frictional force: $[S_4 \lBridge{f} G]$}\label{Sect:dynamicalFriction}

In \S\,\ref{sec:PNBridge} we explained how supplementary forces can be
included in the bridge by adding an extra term to the kick operator.
In this way solving constructing an $N$-body solver that includes the
YORP effect, tidal evolution or post-Newtonian dynamics becomes
relatively straightforward. Here we demonstrate how such an
augmentation to the \bridge\, operator can be used to add dynamical
friction two independent solvers.

\begin{figure}[h!]
\center
\includegraphics[width=0.5\textwidth]{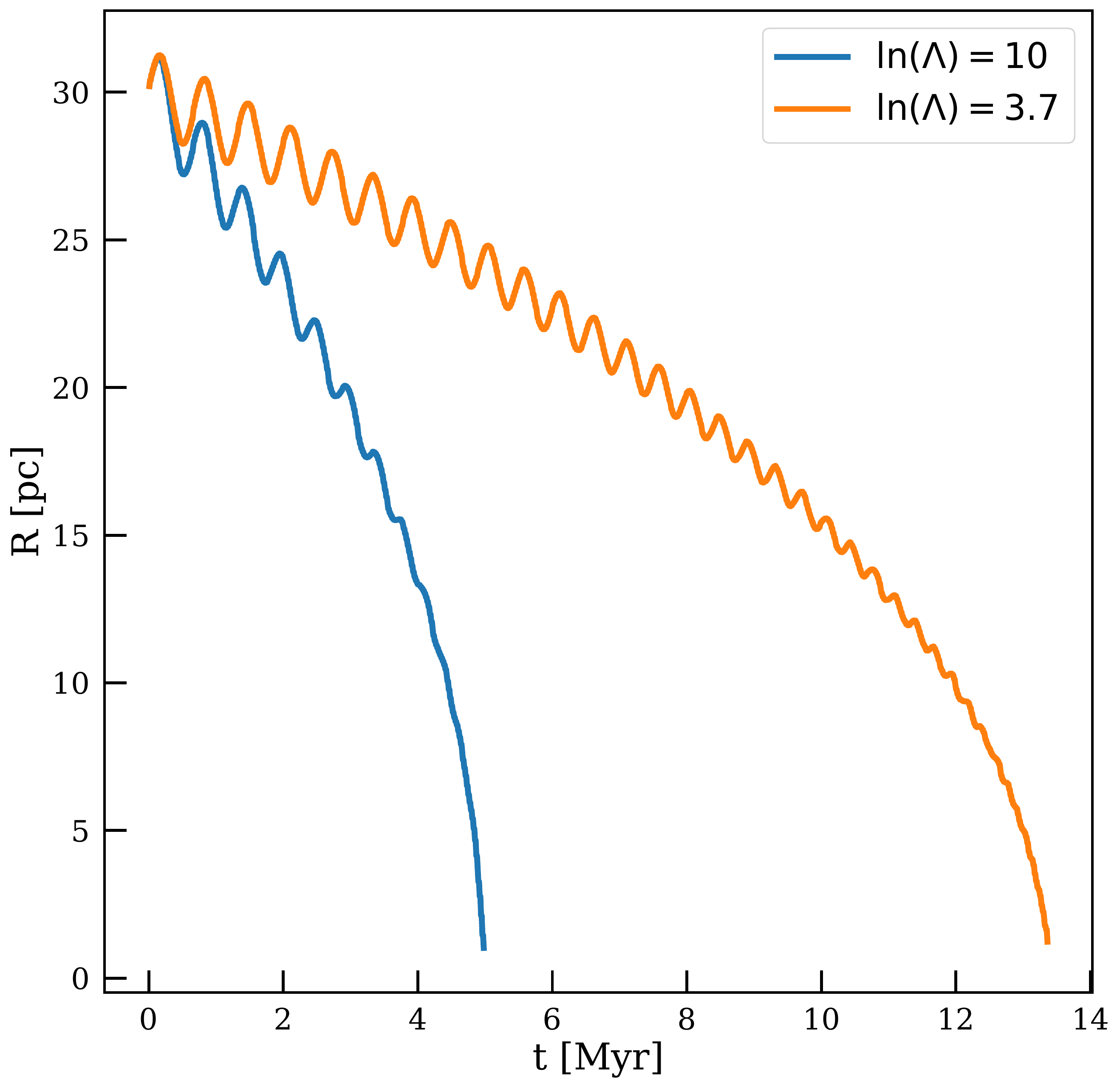}
\caption{ Time evolution of Galactocentric distance $R$ of a star
cluster having an initial mass $M = 10^6$\,M$_\odot$. The
rightmost curve represents a model with $\ln(\Lambda) = 3.7$; the
leftmost curve has $\ln(\Lambda) = 10$. The sinusoidal variations
along the curves result from the slightly eccentric orbits we
adopted. }
\label{fig:bridge_with_friction}
\end{figure}

A star cluster in the Galactic center will feel a frictional force due
to the accumulation of stars in its wake \cite{1987degc.book.....S}.
\cite{2003ApJ...596..314M} used a direct integration technique to
study the time scale on which the Arches star cluster sinks from its
birth location at about 30\,pc from the Galactic center towards the
middle of the Milky Way. It turns out that the time scale depends
quite sensitively on the Coulomb parameter $\ln \Lambda$
\cite{2003MNRAS.344...22S}. In their Fig.6\, they present the results
of several calculations for the distance to the Galactic center as a
function of time for a star cluster of $10^6$\,M$_\odot$. In
fig.\,\ref{fig:bridge_with_friction} we repeat their calculation but
using an external function included in the \bridge\, step, which then
becomes
\begin{equation}
[ S_4 \lBridge{f} G].
\end{equation}
The equation to solve in $f(S,G)$ is Eq.7 of
\cite{2003ApJ...596..314M}.

\begin{figure*}[h!]
\includegraphics[width= 14cm]{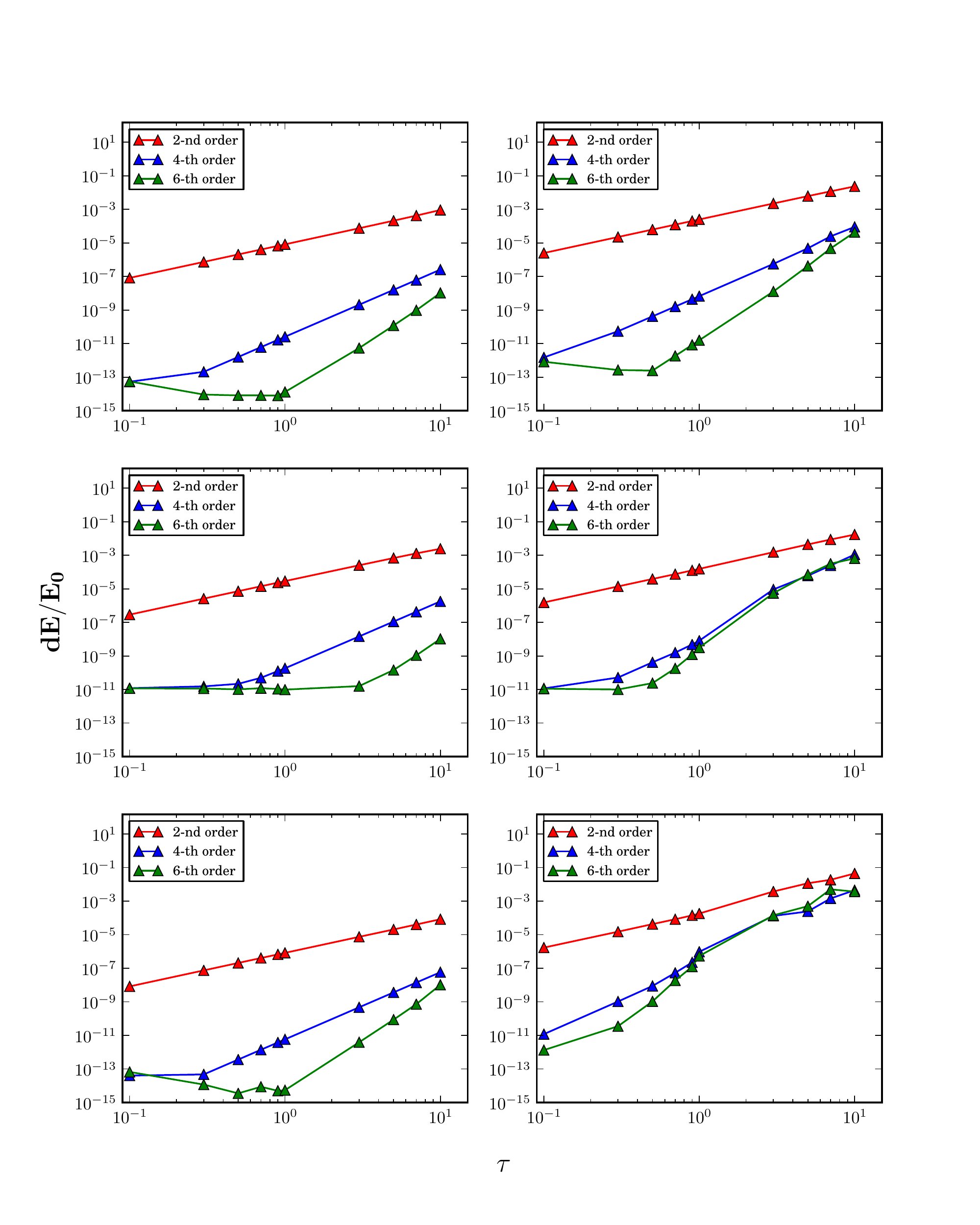}\\
\caption{ Maximum fractional energy error as a function of Bridge time
step of a star moving in a nearly circular orbit ($\epsilon= 0.01$,
left panels) or in an eccentric orbit ($\epsilon=0.5$, right
panels). Presented is the Jocobi energy error, because this is the
quantity conserved in rotating frame. The Milky Way is represented
by: \textit{Top:} A pure axisymmetric potential. \textit{Middle:}
Axisymmetric + bar potential. \textit{Bottom:} Axisymmetric + spiral
arms potential. The red lines correspond to the normal second order
rotating \bridge. The blue and green lines correspond to the
rotating \bridge\ in a fourth and sixth order.
% From top to bottom, the eccentricities of the
% solar-like and eccentric orbits are respectively: $\epsilon= (0.06,
% 0.47); (0.07, 0.5); (0.1, 0.49)$ 
}
\label{fig:energy_error}
\end{figure*}

%\begin{figure}
\begin{codesnippet}[ht]
\centering
\begin{boxedminipage}{9.0cm}
{
\footnotesize
\begin{verbatim}
(1) code1=Hermite()
(2) code1.particles.add_particles(cluster)
(3) code2=DynamicalFriction(Coulomb_logarithm=3.7)
(4) sys=Bridge(timestep=0.1 | units.Myr)
(5) sys.add_system(code1)
(6) sys.add_code(code2)
(7) sys.evolve_model( 10 | units.Myr )
\end{verbatim}
}
\end{boxedminipage}
\caption{ Example usage of an augmented bridge, of the form $[S_4
\lBridge{f} G]$ with an external kick-component to the individual
particles in the $N$-body code.}
\label{fig:PNimplementation}
%%\end{figure}
\end{codesnippet}

In this example, we included the dynamical friction as an additional
drag force in the kick of the bridging step. Alternatively, we could
include a term in the drift-part of the bridge step.

\subsection{Example 4: rotating bridge: $[S_4 \lRBridge{} G_2]$}\label{Sect:rotatingbridge}

To evaluate the accuracy of the Rotating \bridge, we show in
Fig.\,\ref{fig:energy_error} the maximum fractional energy error as a
function of bridge time step of a single star moving in different
galactic potentials representing the Milky Way (axisymmetric, with
bar, and including spiral arms).  Although the microscopic system
contains only a single star, integration of the subsystem is performed
with a $4^{\rm th}$ order scheme. A consequence of only integrating a
single particle warrants that the energy errors presented in
Fig.\,\ref{fig:energy_error} are solely caused by the bridge.

The axisymmetric potential was modeled by taking into account the
parameters of \cite{1991RMxAA..22..255A}. The central bar was modeled
with a Ferrers potential \citep{ferrers} and the spiral arms were
modeled as perturbations of the axisymmetric Galactic potential
following the tight winding approximation
\citep{2014A&A...563A..60A}. Both bar and spiral arms rotate as rigid
bodies with different pattern speeds. For further details on the
Galactic model, we refer the reader to \cite{2015MNRAS.446..823M}.

We also show two different stellar motions. In the left panels, a
single star moves through the Galaxy in a nearly circular orbit with
an eccentricity of $\epsilon=0.01$. In the right panels, the star moves
in an eccentric orbit with $\epsilon=0.5$. We computed both orbits
for ten orbital periods. The circular and eccentric orbits have
orbital periods corresponding to $224$ and $471$ Myr respectively.

We observe that the fourth and sixth-order methods have better energy
conservation compared to the second-order rotating \bridge. In
addition, the energy error is smaller for circular than for eccentric
orbits for a given Galaxy model and \bridge\ time step. This is
expected because the external tidal field is rather constant in a
circular orbit, whereas for $e>0$ it varies. In particular, in
eccentric orbits, a high-order rotating \bridge\ can result in
satisfactory energy conservation at small time steps ($dE/E_0 <
10^{-7}$ at $\tau\leq 1$~Myr), whereas $10^{-3}$ is probably sufficient
\cite{2014ApJ...785L...3P}. The high-order rotating \bridge\ is
suitable for computing the stellar motion in the adopted semi-analytic
galactic background-potential.

\subsection{Example 5: compound hierarchical bridge: the \nemesis\ integrator
$[P_8 \lrBridge{} [S_4 \lBridge{} G_2]]$ }\label{Sect:hierarchical}

\bridge\, operators can become rather elaborate, in particular if they
are hierarchically nested and some operate on different subsets than
others. The \nemesis\ strategy \footnote{The \nemesis\, module is
named after the first application during development in which we
studied a hypothetical companion to the Sun.}, a module in AMUSE,
follows such a strategy.

It is composed of hierarchically nested \bridge\ systems in which the
wide separation in scales allows us to separate the gravitational
forces between the different levels of the hierarchy. The hierarchy
is separated in a macroscopic system (the parent) and the microscopic
system (the children). A parent can have multiple children, but each
child has only one parent. The structure is hierarchical in the sense
that a child can be a parent with multiple children of its own. We
designed such a compounded hierarchical bridge for integrating
multiple planetary systems in star clusters
\cite{2019A&A...624A.120V}, but it is also used for integrating
multiple star clusters in a Galaxy.

In \nemesis, the entire star cluster is separated into one global
structure and any number of sub-structures, which are coupled together
with a cascade of \bridge\, patterns. The global structure is composed
of individual stars and single planets, and integration can be
realized with a direct $N$-body solver. Substructures are treated
individually and, depending on their characteristics, integrated using
a symplectic $N$-body solver.

Each planetary system can now be integrated with its own dedicated
solver, and this is also the case for the star cluster. These
integrators are dynamically created and deleted at run time, depending
on whether or not a subsystem forms or dissolves.

If a planet escapes from its host star, it is picked up by the global
$N$-body integrator whereas the other planets in that same system
continue to be integrated with the dedicated solver. If a planetary
system completely dissolves into individual unbound components, each
star and planet in the subsystem is incorporated into the global
$N$-body code and the subsystem integrator is terminated. If a new
bound subsystem appears, a new (symplectic) $N$-body code to handle the
local dynamics is started and incorporated using a \bridge\, to the
global cluster code.

Asteroids can be taken into account by incorporating an additional
\bridge\, operator with a test-particle integrator, much in the same
way as was introduced in \cite{2014MNRAS.443..355H}. In \nemesis,
stars feel the gravitational force of all other stars and
planets. Planets around a particular star feel the force of each other
and of free-floating planets and the other stars in the cluster,
but not the force of the planets in orbit around other stars.
Asteroids, if included in the calculation, feel the force of all stars
and free-floating planets in the cluster and the local planets, but
not the force of other asteroids or planets around other stars.

\nemesis\ is realized by incorporating four extensions to \bridge,
these include:
\begin{itemize}
\item \emph{topology} \nemesis\ supports two types of systems, (1)
the parent that provides the frame of reference and (2)
children, each of which is a subsystem of its parent. Each child
is represented as a single particle in the parent system.
\item \emph{creation/destruction} A child is created for every close
encounter between two individual particles in the parent. If a
single particle or a child encounters a child, it is absorbed by
the larger child. If a child system contains a single particle,
that particle becomes part of the parent and the particular
integrator is stopped.
\item \emph{exchange} Particles can be exchanged between systems. A
particle that moves too far from a child will be removed from that
system and transferred to the parent. A particle in the parent
that approaches a child system will be incorporated into
the child and removed from the parent.
\item \emph{accounting} A database is maintained to keep track of all
the particles in the parent system and those in the child systems.
This database also keeps track of the running codes.
\end{itemize}

It is rather complicated to express the \nemesis\, module in simple
pseudo-code, but our notation is sufficiently expressive to show the
fundamental structure of this compound hierarchical bridge. Here we
write planets that belong to star $S^i$ as $P(S^i)$. These planets
feel the force of all the other planets in orbit around star $i$, but
not the forces from the planets around other stars $S-S^i$. We first
define the compound system
\begin{equation}
C(P_8, S_4) \equiv [P(S^i)_8 \lrBridge{} S^i_4,
(P(S-S^i))_8 \lrBridge{} (S-S^i)_4]
],
\end{equation}
which we may want to abbreviate to $C(P_8,S_4) \equiv [P(S_4^i)_8,
P(S_4-S_4^i)_8]$. We can subsequently construct a classic bridge
from the compound system $C(P_8,S_4)$ with the galaxy $G_2$ as usual.
\begin{equation}
[C(P_8,S_4) \lBridge{} G_2].
\end{equation}

With this strategy, we create a complex topography of interacting
codes. In its simplest form, {\nemesis} requires 3 different gravity
solvers:
\begin{itemize}
\item [$\bullet$] One code for each of the microscopic systems,
\item [$\bullet$] One code to integrate the macroscopic system,
\item [$\bullet$] One code to calculate all forces and coordinates 
the communication between the other codes.
\end{itemize}

In a more elaborate setup, children themselves could be subdivided
into children to accommodate moons, etc. These sub-children
could be integrated with another method. In principle each child and
each parent can have its dedicated integration method, depending
on the local requirements. It is even possible to built-up multiple
{\nemesis} modules hierarchically to make an even more complicated
compound.

\subsubsection{Testing {\nemesis}}

To illustrate the working of {\nemesis} we present the results of a
calculation in which we integrated the multiple hierarchical
configuration for simulating a star cluster with planetary systems,
but without a galactic background. This is only the
$C(P_8,S_4)$ component of the compound solver.

We test the setups by integrating an isolated system of five planets
in orbit around a 1\,M$_\odot$ star and a perturbing star in a wide
orbit. The planetary system is generated using the oligarchic growth
model \citep{1998Icar..131..171K} for a 1\,M$_\odot$ star with a
400\,au disk of 0.1\,M$_\odot$.  The outer star of 1M$_\odot$ has a
semi-major axis of $1500$\,au with $0.5$ eccentricity and an
inclination of $90^\circ$.

The simulations were performed using {\nemesis and integrating all objects in a single $N$-body code. The {\nemesis} method was
constructed using one code for the planetary system and one code for
the center of mass of the planetary system and the orbiting secondary
star. For both integrators in \nemesis, we adopted the eighth order
symplectic integrator in {\tt Huayno} \citep{2012NewA...17..711P}.
This system was integrated using a nemesis time step of $dt_{\tt
Nemesis} = 100$\,yr. The calculation with \nemesis\ method took
about ten minutes on a 3.6\,GHz core i7-4790 based workstation: the
Hermite scheme (green curve) took 18 hours on the same machine. The
resulting evolution of the energy error is presented in In
Fig.\,\ref{fig:simplecticity}.

\begin{figure}[htb]
\centering
\includegraphics[width=0.8\textwidth]{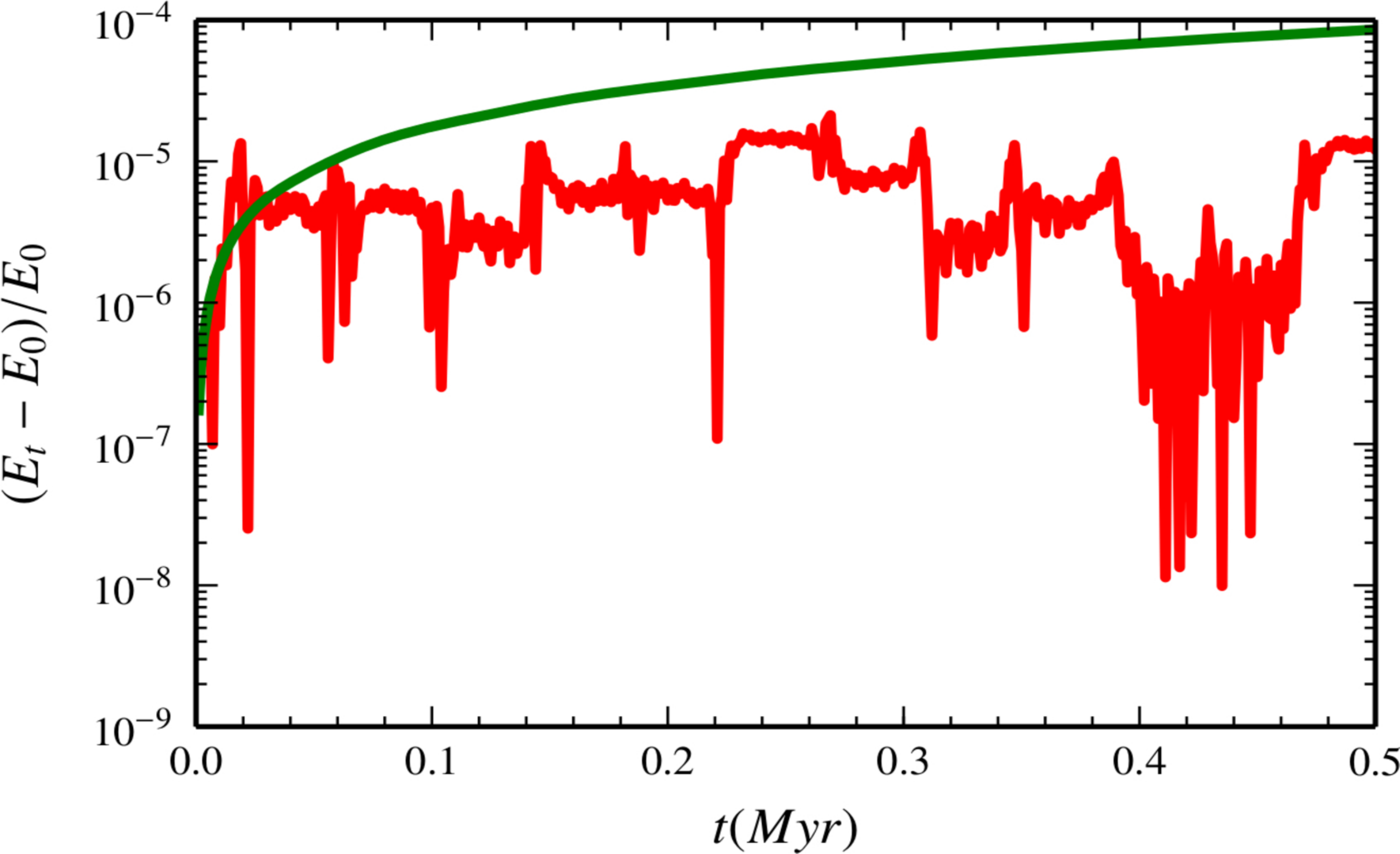}
\caption{Total energy error as a function of time for a validation
simulation consisting of a wide binary of which one star is orbited
by 5 planets (see \cite{2019A&A...624A.120V}). The energy error of
the \nemesis\ method (in red) is compared with the results obtained
using a single $4^\textrm{th}$-order Hermite code for all particles (smooth
green curve). The time evolution of the energy error is more
erratic in the \nemesis\ method because of the close interactions of
the orbiting star. The overall error, however, remains rather
constant over a long timescale, whereas for the Hermite method the
energy error is smoother but gradually grows with time,
characteristic for a non-symplectic integrator.}
\label{fig:simplecticity}
\end{figure}

The energy error in the Hermite (smooth green curve) grows monotonically,
which is the typical response for a non-symplectic integrator, such as
the adopted Hermite scheme. The evolution of the energy error in the
hybrid integrator does not grow on a secular timescale. The evolution
of the energy error is rather erratic with sharp peaks to low values
as well as high values but stays stable overall. The secular growth of
\nemesis\ is much smaller than the single Hermite integrator.

\section{Conclusions}
\label{sec:discussion_b}

We introduce a coupling strategy allowing numerical integrators to
co-operate and interact without the requirement of making changes to
the underlying codes. We show how this non-intrusive method is
symplectic, it can be expanded to higher (even) orders, and how it can
be augmented to include other terms in the operator. The scheme is
based on the fundamental work of \citep{2007PASJ...59.1095F}, which by
introducing \bridge\, provides a powerful basis for building complex
integrators while keeping the underlying codes dedicated, small and
clean.

The symplectic behavior of \bridge\, is preserved when symplectic
codes are coupled. When non-symplectic codes are coupled the
symplectic quality of the compound solver is lost also. The
consequential symplectic growth in the integration error can be
compensated in part by adopting a higher-order integrator for the
coupling strategy. The same strategy can be used when the individual
coupled codes exhibit higher-order behavior.

We demonstrated how the scheme can be extended to higher-order, as
well as to multiple interacting systems and deeper hierarchies of
nested subsystems. Hierarchically coupled bridges provide a high-order
symplectic coupling between two or more subsystems. The formulation is
independent of the actual implementation of the evolution operators.

An additional advantage of the non-intrusive bridge
presented here is the possibility to introduce variations in the
spatial or temporal coordinate system and by additional forces in the
interface. This allows us to introduce additional physical processes,
such as post-Newtonian terms or dynamical friction without changes to
the underlying codes. The main advantage of such a procedure is its
ability to define a general high-level coupling strategy, which again
allows underlying solvers to remain small and simple.

We tested the method on self-gravitating astrophysical systems in
which we coupled Newtonian force evaluators via a (symplectic and
non-symplectic) direct $N$-body and hierarchical $N$-body (tree)
codes. We confirm that energy conservation has the right order for the
second (eq.\,\ref{eq:B2}), fourth (eq.\,\ref{eq:B4}), sixth
(eq.\,\ref{eq:B6}), eighth (eq.\,\ref{eq:B8}) and tenth
(eq.\,\ref{eq:B10}) order coupling strategies, up to machine precision
($dE/E \simeq 10^{-15}$), if the limitations regarding the component
integrators are satisfied. Within \amuse\, up to tenth order methods
are available. The method is hierarchical and the same expansion can
be repeated to construct more complicated bridge structure.

We also present an implementation for integration in a rotating frame
of reference. The formulation follows a similar splitting argument,
where a formally symplectic scheme follows if we use canonical
coordinates. A non-canonical formulation is somewhat easier to combine
with self-interacting systems. The difference with the classic
\bridge\ is that the kick operator affecting the interactions between
the different subsystems must be adapted. Note that the evolution
between kicks has not changed form, and the same integrator can be
used for both operations. In the non-canonical case the resulting
integrator is not formally symplectic (and this manifests itself as a
net drift in energy), but for practical applications (e.g. a stellar
cluster in a galaxy potential) the resulting drift is, due to the
relatively low number of orbits, negligible in particular if combined
with a higher-order integration scheme.

We finally demonstrate how multiple \bridge\ operators can be combined
to construct a hierarchical compound system in which a subset of
particles selectively feel the force of some other subset. This
selective hierarchical nesting of bridge operators preserves the
symplectic quality of the individual integrators. We call this method
\nemesis\, and it is used to study planetary systems in star clusters.

The operators are implemented in the public \amuse\ and
\omuse\ frameworks. Together with a wide variety of different
numerical integrators, \amuse\, provides the ready-made building
blocks for combining them to a tailored application to study
astrophysical phenomena.

\section*{Acknowledgements}
We thank Jeroen B\'edorf, Tjarda Boekholt, Michiko Fujii, Guilherme
Gon\c calves Ferrari, Adrian Hamers, Lucie J\`{\i}lkov\'a and Eugene
Vasiliev for enriching discussions. In this work we used the
following packages: AMUSE
\citep{2011ascl.soft07007P,portegies_zwart_simon_2018_1443252}, Galpy
\cite{2014ascl.soft11008B}, Hermite0 \citep{2014DDA....4530301M},
Huayno \citep{2012NewA...17..711P}, matplotlib
\citep{2007CSE.....9...90H}, numpy \citep{Oliphant2006ANumPy}, and
Python \citep{vanRossum:1995:EEP}. This work was supported by the
Netherlands Research School for Astronomy (NOVA), NWO (grant \#
621.016.701 [LGM-II]) and by the European Union's Horizon 2020
research and innovation program under grant agreement No 671564
(COMPAT project). This work was in part done at the Canadian
Institute for Theoretical Astronomy and SPZ is grateful for their
support, in particular to Norm Murray who made this possible.

%\input /home/spz/Latex/lib/bib/references

%\bibliographystyle{/home/spz/Latex/lib/styles/elsevier/elsarticle-num} 
%\bibliography{references} 

\vfill
\newpage

\section*{Appendix A}\label{Sect:AppendixA}

Overview of the coefficients used for high-order symplectic
integration \cite{1990PhLA..150..262Y,Hairer2005} used in
section\,\ref{sec:Highorderbridge}.

The various integrators are indicated with S\#M\# (or equivalently
${\cal B}_\textrm{S}^\textrm{M}$). Here the first \# indicates the
splitting strategy; the symmetric composition {\tt S} and the number
of evaluations per bridge step {\tt M}. The equations associated with
the integration constants listed in table\,\ref{Table:Coefficients}.

\begin{eqnarray}
  {\cal B}_2 &=& K(u_0\tau)D(v_0\tau)K(u_0\tau) \label{eq:B2} \\
  {\cal B}_4^4 &=& K(u_0\tau)D(v_0\tau)
                 K(u_1\tau)D(v_1\tau)
                 K(u_2\tau)D(v_1\tau)
                 K(u_1\tau)D(v_0\tau)
                 K(u_0\tau) \label{eq:B4}\\ 
  {\cal B}_4^5 &=& K(u_0\tau)D(v_0\tau)
                 K(u_1\tau)D(v_1\tau)
                 K(u_2\tau)D(v_2\tau)
                 K(u_2\tau)D(v_1\tau)
                 K(u_1\tau)D(v_0\tau)
                 K(u_0\tau)          \\
  {\cal B}_4^6 &=& K(u_0\tau)D(v_0\tau)
                   K(u_1\tau)D(v_1\tau)
                   K(u_2\tau)D(v_2\tau)
                   K(u_3\tau)D(v_2\tau)
                   K(u_2\tau)D(v_1\tau)\nonumber \\
               & & K(u_1\tau)D(v_0\tau)
                   K(u_0\tau) \\
  {\cal B}_6^{11} &=&  K(w_0\tau/2)  D(w_0\tau)
                      K((w_0+w_1)\tau/2)  D(w_1\tau)
                      K((w_1+w_2)\tau/2)  D(w_2\tau) \nonumber \\
                  & & K((w_2+w_3)\tau/2)  D(w_3\tau) 
                      K((w_3+w_4)\tau/2)  D(w_4\tau)
                      K((w_4+w_5)\tau/2)  D(w_5\tau) \nonumber \\
                  & & K((w_5+w_4)\tau/2)  D(w_4\tau)
                      K((w_4+w_3)\tau/2)  D(w_3\tau)
                      K((w_3+w_2)\tau/2)  D(w_2\tau)  \nonumber \\
                  & & K((w_2+w_1)\tau/2)  D(w_1\tau)
                      K((w_1+w_0)\tau/2)  D(w_0\tau)
                      K((w_0)\tau/2)  \label{eq:B6} \\
  {\cal B}_6^{13} &=&  K(w_0\tau/2)  D(w_0\tau)
                      K((w_0+w_1)\tau/2)  D(w_1\tau)
                      K((w_1+w_2)\tau/2)  D(w_2\tau)\nonumber \\
                 & &  K((w_2+w_3)\tau/2)  D(w_3\tau)
                      K((w_3+w_4)\tau/2)  D(w_4\tau)
                      K((w_4+w_5)\tau/2)  D(w_5\tau)\nonumber \\
                 & &  K((w_5+w_6)\tau/2)  D(w_6\tau)
                      K((w_6+w_5)\tau/2)  D(w_5\tau)
                      K((w_5+w_4)\tau/2)  D(w_4\tau)\nonumber \\
                 & &  K((w_4+w_3)\tau/2)  D(w_3\tau)
                      K((w_3+w_2)\tau/2)  D(w_2\tau)
                      K((w_2+w_1)\tau/2)  D(w_1\tau)\nonumber \\
                 & &  K((w_1+w_0)\tau/2)  D(w_0\tau)
                      K(w_0\tau/2) \\
  {\cal B}_8^{21} &=&  K(w_0\tau/2)  D(w_0\tau)
                      K((w_0+w_1)\tau/2)  D(w_1\tau)
                      K((w_1+w_2)\tau/2)  D(w_2\tau) \nonumber \\
                 & &  K((w_2+w_3)\tau/2)  D(w_3\tau)
                      K((w_3+w_4)\tau/2)  D(w_4\tau)
                      K((w_4+w_5)\tau/2)  D(w_5\tau)\nonumber \\
                 &&     K((w_5+w_6)\tau/2)  D(w_6\tau)
                      K((w_6+w_7)\tau/2)  D(w_7\tau)
                      K((w_7+w_8)\tau/2)  D(w_8\tau)\nonumber \\
                 &&     K((w_8+w_9)\tau/2)  D(w_9\tau)
                      K((w_9+w_{10})\tau/2)  D(w_{10}\tau)
                      K((w_{10}+w_9)\tau/2)  D(w_9\tau)\nonumber \\
                 &&      K((w_9+w_8)\tau/2)  D(w_8\tau)
                      K((w_8+w_7)\tau/2)  D(w_7\tau)
                      K((w_7+w_6)\tau/2)  D(w_6\tau)\nonumber \\
                  &&    K((w_6+w_5)\tau/2)  D(w_5\tau)
                      K((w_5+w_4)\tau/2)  D(w_4\tau)
                      K((w_4+w_3)\tau/2)  D(w_3\tau)\nonumber \\
                  &&    K((w_3+w_2)\tau/2)  D(w_2\tau)
                      K((w_2+w_1)\tau/2)  D(w_1\tau)
                      K((w_1+w_0)\tau/2)  D(w_0\tau)\nonumber \\
                  &&    K(w_0\tau/2) \label{eq:B8} \\
  {\cal B}_{10}^{35} &=&  K(w_0\tau/2)  D(w_0\tau)
                         K((w_0+w_1)\tau/2)  D(w_1\tau)
                         K((w_1+w_2)\tau/2)  D(w_2\tau)\nonumber \\
                    & &  K((w_2+w_3)\tau/2)  D(w_3\tau)
                         K((w_3+w_4)\tau/2)  D(w_4\tau)
                         K((w_4+w_5)\tau/2)  D(w_5\tau)\nonumber \\
                    & &     K((w_5+w_6)\tau/2)  D(w_6\tau)
                         K((w_6+w_7)\tau/2)  D(w_7\tau)
                         K((w_7+w_8)\tau/2)  D(w_8\tau)\nonumber \\
                    & &  K((w_8+w_9)\tau/2)  D(w_9\tau)
                         K((w_9+w_{10})\tau/2)  D(w_{10}\tau)
                         K((w_{10}+w_{11})\tau/2)  D(w_{11}\tau)\nonumber \\
                   & &   K((w_{11}+w_{12})\tau/2)  D(w_{12}\tau)
                         K((w_{12}+w_{13})\tau/2)  D(w_{13}\tau)
                         K((w_{13}+w_{14})\tau/2)  D(w_{14}\tau)\nonumber \\
                   & &      K((w_{14}+w_{15})\tau/2)  D(w_{15}\tau)
                         K((w_{15}+w_{16})\tau/2)  D(w_{16}\tau)
                         K((w_{16}+w_{17})\tau/2)  D(w_{17}\tau)\nonumber \\
                    & &     K((w_{15}+w_{16})\tau/2)  D(w_{16}\tau)
                         K((w_{14}+w_{15})\tau/2)  D(w_{15}\tau)
                         K((w_{13}+w_{14})\tau/2)  D(w_{14}\tau)\nonumber \\
                     & &    K((w_{12}+w_{13})\tau/2)  D(w_{13}\tau)
                         K((w_{11}+w_{12})\tau/2)  D(w_{12}\tau)
                         K((w_{10}+w_{11})\tau/2)  D(w_{11}\tau)\nonumber \\
                     & &    K((w_{10}+w_9)\tau/2)  D(w_9\tau)
                         K((w_9+w_8)\tau/2)  D(w_8\tau)
                         K((w_8+w_7)\tau/2)  D(w_7\tau)\nonumber \\
                     & &    K((w_7+w_6)\tau/2)  D(w_6\tau)
                         K((w_6+w_5)\tau/2)  D(w_5\tau)
                         K((w_5+w_4)\tau/2)  D(w_4\tau)\nonumber \\
                     & &    K((w_4+w_3)\tau/2)  D(w_3\tau)
                         K((w_3+w_2)\tau/2)  D(w_2\tau)
                         K((w_2+w_1)\tau/2)  D(w_1\tau)\nonumber \\
                      & &   K((w_1+w_0)\tau/2)  D(w_0\tau)
                         K(w_0\tau/2) \label{eq:B10} 
\end{eqnarray}

\begin{table}
%% \centering
 \caption{Coefficeints for the higher-order symplectic bridge
   implementations, see \cite{1990PhLA..150..262Y,Hairer2005}.}
 \label{Table:Coefficients}
\begin{tabular}{llll}
 \hline  
 \hline
 \multicolumn{2}{l}{{\tt S}2{\tt M}2}\\
$u_0$ & $1/2$ \\
$v_0$ & $1$ \\
 \hline
 \multicolumn{2}{l}{{\tt S}4{\tt M}4} &  \multicolumn{2}{l}{{\tt S}4{\tt M}5}\\
$u_0$ & $(642 + 471^{1/2})/3924$       & $(14 - 19^{0.5})/108$\\
$u_1$ & $121 (12- 471^{1/2})/3924$     & $(20 - 7 \cdot 19^{0.5})/108$\\
$u_2$ & $1 - 2 (u_0 + u_1)$           & $1/2-(u_0+u_1)$\\
$v_0$ & $6/11$                        & $2/5$\\
$v_1$ & $1/2 - v_0$                   & $-1/10$\\
$v_2$ &                               & $1-2(v_0+v_1)$ \\ 
 \hline
 \multicolumn{2}{l}{{\tt S}4{\tt M}6}\\
$u_0$ & $0.0792036964311957$\\
$u_1$ & $0.353172906049774$\\
$u_2$ & $-0.0420650803577195$\\
$u_3$ & $1. - 2(u_0 + u_1 + u_2)$\\
$v_0$ & $0.209515106613362$\\
$v_1$ & $-0.143851773179818$\\
$v_2$ & $0.5 - v_0 - v_1$\\
\hline  
\multicolumn{2}{l}{{\tt S}6{\tt M}11} & \multicolumn{2}{l}{{\tt S}6{\tt M}13}\\
$w_0$& 0.21375583945878254555518066964857 &0.13861930854051695245808013042625\\
$w_1$& 0.18329381407425713911385974425217 &0.13346562851074760407046858832209\\
$w_2$& 0.17692819473098943794898811709929 &0.13070531011449225190542755785015\\
$w_3$&-0.44329082681170215849622829626258 &0.12961893756907034772505366537091\\
$w_4$& 0.11728560432865935385403585669136 &-0.35000324893920896516170830911323\\
$w_5$& 0.50405474843802736404832781714239 &0.11805530653002387170273438954049\\
$w_6$&                                    &0.39907751534871587459988795520665\\
\hline  
 \multicolumn{2}{l}{{\tt S}8{\tt M}21} &\multicolumn{2}{l}{{\tt S}10{\tt M}35}\\
$w_0$ & 0.10647728984550031823931967854896& 0.078795722521686419263907679337684\\
$w_1$ & 0.10837408645835726397433410591546& 0.31309610341510852776481247192647\\ 
$w_2$ & 0.35337821052654342419534541324080& 0.027918383235078066109520273275299\\
$w_3$ &-0.23341414023165082198780281128319&-0.22959284159390709415121339679655\\ 
$w_4$ &-0.24445266791528841269462171413216& 0.13096206107716486317465685927961\\ 
$w_5$ & 0.11317848435755633314700952515599&-0.26973340565451071434460973222411\\ 
$w_6$ & 0.11892905625000350062692972283951& 0.074973343155891435666137105641410\\
$w_7$ & 0.12603912321825988140305670268365& 0.11199342399981020488957508073640\\ 
$w_8$ & 0.12581718736176041804392391641587& 0.36613344954622675119314812353150\\ 
$w_9$ & 0.11699135019217642180722881433533&-0.39910563013603589787862981058340\\ 
$w_{10}$&-0.38263596012643665350944670744040& 0.10308739852747107731580277001372\\
$w_{11}$ &&   0.41143087395589023782070411897608 \\
$w_{12}$ &&  -0.0048663605831352617621956593099771 \\
$w_{13}$ &&  -0.39203335370863990644808193642610 \\
$w_{14}$ &&   0.051942502962449647037182904015976 \\
$w_{15}$ &&   0.050665090759924496335874344156866 \\
$w_{16}$ &&   0.049674370639729879054568800279461 \\
$w_{17}$ &&   0.049317735759594537917680008339338 \\
\hline   
\hline  
\end{tabular}
\end{table}

\end{document}